\documentclass[12pt,dvipdfmx]{article}

\usepackage{amsmath,amssymb,amsfonts,graphics,graphicx,amscd,amsfonts,epsfig,color}
\usepackage{subfig}
\newcommand {\beq} {\begin{equation}}
\newcommand {\eeq} {\end{equation}}
\newcommand {\beqa}{\begin{eqnarray}}
\newcommand {\eeqa}{\end{eqnarray}}
\newcommand {\nn} {\nonumber}
\newcommand {\del} {\partial}
\newcommand {\tr}{{\rm tr\,}}

\newcommand {\ee}{\mbox{e}}

\newcommand{\bbR}{{\mathbb R}}
\newcommand{\bbC}{{\mathbb C}}
\makeatletter
    
    \@addtoreset{equation}{section}
  \makeatother
\allowdisplaybreaks[4]

\date{}
\begin{document}

\begin{flushright} 
%\today\\
% WIS/06/09-MAY-DPP\\ 
KEK-TH-1911
\end{flushright} 

\vspace{0.1cm}

\begin{center}
{\LARGE The argument for justification of the complex Langevin method and
the condition for correct convergence}
%{\LARGE The argument for justification of the complex Langevin method revisited}
%%   {\LARGE
%% Subtlety in taking the zero step-size limit 
%% in the complex Langevin method for complex-action systems
%% %A note on justification of the gauge cooling 
%% %Subtlety in taking the continuum Langevin time limit 
%% %%
%% %as the real obstacle 
%% %
%% %% A refined argument for justifying 
%% %% the complex Langevin method and
%% %% the necessary and sufficient condition
%% %%
%% % for justification
%% %On the necessary and sufficient condition for justifying 
%% %the complex Langevin method
%% %
%% %% \\
%% %% [0.2cm]
%% %% with the gauge cooling procedure
%% %
%% %% \\[0.2cm]
%% %% to the complex action problem
%% %
%% %\\[0.2cm]
%% %for D0-branes at Finite Temperature and Finite $N$
%%   }
\end{center}
\vspace{0.1cm}
\vspace{0.1cm}
\begin{center}

         Keitaro N{\sc agata}$^{a}$\footnote
          {
 E-mail address : knagata@post.kek.jp},  
         Jun N{\sc ishimura}$^{ab}$\footnote
          {
 E-mail address : jnishi@post.kek.jp} 
and
         Shinji S{\sc himasaki}$^{ac}$\footnote
          {
 E-mail address : shinji.shimasaki@keio.jp} 

\vspace{0.5cm}

$^a${\it KEK Theory Center, 
High Energy Accelerator Research Organization,\\
1-1 Oho, Tsukuba, Ibaraki 305-0801, Japan}

$^b${\it Graduate University for Advanced Studies (SOKENDAI),\\
1-1 Oho, Tsukuba, Ibaraki 305-0801, Japan} 

$^c${\it Research and Education Center for Natural Sciences, Keio University,\\
Hiyoshi 4-1-1, Yokohama, Kanagawa 223-8521, Japan}

\end{center}
%\newpage

\vspace{1.5cm}

\begin{center}
  {\bf abstract}
\end{center}

\noindent The complex Langevin method is a promising approach
to the complex-action problem
based on a fictitious time evolution of 
complexified dynamical variables under the influence 
of a Gaussian noise.
%Unlike the stochastic quantization for a real action,
%conventional Monte Carlo methods,
Although it is known to have a restricted range of applicability,
the use of 
%so-called 
gauge cooling
%in models with large symmetry
%was shown to make 
made it applicable to various interesting cases
including finite density QCD in certain parameter regions.
%which were not accessible before.
In this paper we revisit the argument for justification of the method.
In particular, we point out a subtlety 
in the use of time-evolved observables, 
which play a crucial role in the previous argument.
This requires that
the probability of the drift term should fall off
exponentially or faster at large magnitude.
We argue that this is actually a necessary and sufficient condition
for the method to be justified.
%We also point out that the integration by parts, which was considered
%to be the cause of the problem in the previous argument,
%% Starting with a finite Langevin step-size $\epsilon$,
%% we find that the $\epsilon\rightarrow 0$ limit is actually subtle,
%% although the previous argument used a continuous Langevin time from the outset.
% in a unified way.
Using two simple examples, we show that our condition
%provides the necessary and sufficient condition for the method to work,
%which 
%can be used to judge 
tells us
%is useful in 
clearly whether the results obtained by the method
are trustable or not.
We also discuss a new possibility for the gauge cooling, which
can reduce the magnitude of the drift term directly.
% in practical applications.
%it can be checked easily in practical applications.
%it is useful in practical applications 
%justification explicitly in a simple case with a pole in the drift term.
%% Our refined argument
%% also enables
%% us to formulate the necessary and sufficient
%% condition for 
%% justification
%% in such a way that it can be tested explicitly by numerical computations.
%% which is tested explicitly in a simple case with a pole in the drift term.
%in such a way that it can be tested explicitly by numerical computations.
%
%This condition includes the necessary condition proposed before,
%and we test it in a simple example with a singular drift term.
%
%% We test this condition explicitly in a simple model
%% and demonstrate that it can indeed
%% be used to tell 
%% whether the results 
%% obtained by the method 
%% are trustable or not.
%
% without knowing the correct results beforehand.
%This is in contrast to the situation
%with the continuous time, where the interchangability
%of the description can be lost due to the unjustifiable
%integration by parts used in the argument.

%%%%%%%%%%%%%%%%%%%%%%%%%%%%%%%%%%%%%%%%%%%%%%%%%%%%%%%%%%%%%%%%%%%%%%
%%%%%%%%%%%%%%%%%%%%%%%%%%%%%%%%%%%%%%%%%%%%%%%%%%%%%%%%%%%%%%%%%%%%%%
%%%%%%%%%%%%%%%%%%%%%%%%%%%%%%%%%%%%%%%%%%%%%%%%%%%%%%%%%%%%%%%%%%%%%%
\newpage

%%%%%%%%%%%%%%%%%%%%%%%%%%%%%%%%%%%%%%%%%%%%%%%%%%%%%%%%%%%%%%%%%%%%%%
%%%%%%%%%%%%%%%%%%%%%%%%%%%%%%%%%%%%%%%%%%%%%%%%%%%%%%%%%%%%%%%%%%%%%%
%%%%%%%%%%%%%%%%%%%%%%%%%%%%%%%%%%%%%%%%%%%%%%%%%%%%%%%%%%%%%%%%%%%%%%
\section{Introduction}
%\hspace{0.51cm}
%%%%%%%%%%%%%%%%%%%%%%%%%%%%%%%%%%%%%%%%%%%%%%%%%%%%%%%%%%%%%%%%%%%%%%
%%%%%%%%%%%%%%%%%%%%%%%%%%%%%%%%%%%%%%%%%%%%%%%%%%%%%%%%%%%%%%%%%%%%%%
%%%%%%%%%%%%%%%%%%%%%%%%%%%%%%%%%%%%%%%%%%%%%%%%%%%%%%%%%%%%%%%%%%%%%%

Solving the negative sign problem or more generally
the complex-action problem is one of the most important challenges
in computational science.
This is a problem that occurs in an attempt to apply the
idea of importance sampling to multiple integration 
with a weight which 
%an integrand which 
%is not positive semi-definite.
fluctuates
in sign or in its complex phase.
%overa huge number of variables when the integrand fluctuates
%in sign or in its complex phase.
The complex Langevin method (CLM)
\cite{Parisi:1984cs,Klauder:1983sp}
is a promising approach,
%to this problem
which can be applied to a variety of models with a complex weight
albeit not all of them.
%statistical systems.
For instance, it has been applied successfully 
to finite density QCD either with heavy quarks \cite{Aarts:2014bwa}
or in the deconfined phase \cite{Sexty:2013ica}
using a new technique called gauge cooling \cite{Seiler:2012wz}.
Whether it is applicable also in the case with light quarks
and in the confined phase is 
%an important open question, which is 
one of the hottest topics 
in this field \cite{Mollgaard:2013qra,Mollgaard:2014mga,%
Fodor:2015doa,Sinclair:2015kva,Nagata:2016alq}.
%% which is now being discussed 
%% intensively \cite{Mollgaard:2013qra,Mollgaard:2014mga,%
%% Fodor:2015doa,Sinclair:2015kva,NNS}.
% in the particle physics
%community \cite{Mollgaard:2013qra,Mollgaard:2014mga,Sinclair:2015kva}.

The CLM
%complex Langevin method 
may be viewed as a generalization
of the stochastic quantization \cite{Parisi:1980ys},
%for the case with a positive semi-definite weight.
which generates dynamical variables
with a given probability 
by solving the Langevin equation
that describes a fictitious time evolution of those variables
under the influence of a Gaussian noise.
(See ref.~\cite{Damgaard:1987rr} for a comprehensive review.)
%sequentially.
%corresponding to the positive definite weight
When one applies this idea to the calculation of expectation
values of observables with a complex weight,
%the weight becomes complex, 
one necessarily has to complexify
the dynamical variables
%in solving the corresponding Langevin equation.
due to the complex drift term, which is derived from the complex weight.
Correspondingly, the drift term and the observables 
should be extended to holomorphic functions of the complexified variables
by analytic continuation. Then by measuring 
the observables for the complexified variables
generated by the Langevin process
% and taking the average in the long time limit,
and calculating their expectation values at sufficiently late times,
%which are also
%extended to holomorphic functions of the complexified variables,
one can obtain the expectation values of 
the observables for the original real variables with the complex weight.

It has been known for a long time that this method does not always work.
%For instance, 
Typically, the complex Langevin process reaches thermal equilibrium
without any problem,
but the results for the expectation values
obtained in the way mentioned above turn out to be simply wrong in some cases.
%fails in some cases.
The reason for the failure was discussed
in refs.~\cite{Aarts:2009uq,Aarts:2011ax}
starting from the complex Langevin equation 
with a continuous Langevin time.
There, it was found that 
a subtlety 
%in justifying the method comes from
exists in
the integration by parts used in 
translating the time evolution of the probability distribution
of the complexified variables into that of the observables.
In order for the integration by parts to be valid,
%the boundary terms should vanish.
%This requires that 
the probability distribution
of the complexified variables should have appropriate 
asymptotic behaviors.
By now, the following two conditions are recognized.
\begin{enumerate}
\item The probability distribution should be
suppressed strongly enough when 
the complexified variables take large values \cite{Aarts:2009uq,Aarts:2011ax}.
%% Typically, it is the imaginary parts of the complexified 
%% dynamical variables that may cause a problem.
Typically, this becomes a problem when the complexified
variables make long excursions in the imaginary directions 
during the Langevin simulation.
\item The drift term 
%in the complex Langevin equation
can have singularities
while it is otherwise a holomorphic function of the complexified variables.
In that case, the probability distribution should be suppressed strongly enough
near the singularities \cite{Nishimura:2015pba}.
\end{enumerate}

In fact, both these conditions are relevant
in applying the CLM to finite density QCD.
The condition 1 is an issue
because the link variable, upon complexification,
becomes an ${\rm SL}(3,\bbC)$ matrix, which forms
a noncompact manifold. 
Here, the idea of gauge cooling turned out to be useful \cite{Seiler:2012wz}.
It is based on the fact that
the ${\rm SU}(3)$ gauge symmetry of the action and the observables
is enhanced to the ${\rm SL}(3,\bbC)$ gauge symmetry
upon complexification of the dynamical variables.
One can actually make a complexified gauge transformation
after each Langevin step in such a way that the link variables
stay close to the original ${\rm SU}(3)$ manifold
during the Langevin simulation.
Using this technique, the CLM became applicable to finite density
QCD in the heavy dense limit \cite{Seiler:2012wz,Aarts:2013uxa,Aarts:2016qrv}
and in the deconfined phase \cite{Sexty:2013ica,Fodor:2015doa}.
%certain parameter regions

The condition 2 is also an issue in finite density QCD
because the drift term 
${\rm Tr}[ (D+m)^{-1} \partial (D+m)]$, 
which comes from the fermion determinant,
has singularities corresponding to the appearance of 
zero eigenvalues of the Dirac operator $D+m$.
This becomes a problem
% in the confined phase 
at low temperature
when the mass is small
as demonstrated clearly in the chiral 
Random Matrix Theory (cRMT) \cite{Mollgaard:2013qra}.
In the case of cRMT, 
changing the integration variables in the original path
integral to the polar 
coordinates
%Cartesian coordinates for the matrix elements in the original path integral 
was shown to solve the problem \cite{Mollgaard:2014mga}.
This is possible because the change of variables
in the original path integral leads to 
an inequivalent complex Langevin 
process\footnote{The reason why it works in this case is rather trivial, 
though. After complexification of the polar coordinates, 
the chemical potential $\mu$ can be absorbed
by shifting the imaginary part of the angular variables. 
Thus the complex Langevin equation reduces to that for $\mu=0$,
which does not have any problem. 
Also it is not obvious how one can extend this idea to finite density QCD.}.
In our previous publication \cite{Nagata:2016alq}, we proposed that 
it should be possible to solve this problem also
%should also be solved 
by the gauge cooling with different criteria for choosing
the complexified gauge transformation. 
The results for the cRMT look promising.

%As the history reveals, understanding the problem is always important

%% The usefulness of the CLM, if naively implemented,
%% would be quite restricted because of the above conditions.
%% However, there are various ways in which one can extend the method
%% in such a way that the conditions are satisfied.

The argument for justification of the CLM
given in refs.~\cite{Aarts:2009uq,Aarts:2011ax}
has been extended to the case including 
the gauge cooling procedure recently \cite{Nagata:2015uga}.
This settles down various skepticism 
on the validity of gauge cooling.
For instance, the gauge cooling uses 
the complexified gauge symmetry, which is not
respected by the noise term in the complex Langevin equation.
Despite such issues,
the argument for justification goes through 
as is shown explicitly in ref.~\cite{Nagata:2015uga}.
%The explicit demonstration given in ref.~\cite{Nagata:2015uga}
%shows that this is 

In this paper, we revisit the argument for justification of the CLM
with or without the gauge cooling procedure.
In particular,
we point out a subtlety in the use of time-evolved observables,
which play a crucial role in the argument.
In the previous argument, it was assumed implicitly that
time-evolved observables can be used for infinitely long time.
We argue that this assumption is too strong.
%%We show that this is a too strong assumption, which 
%%is not always satisfied.
In fact, we only need to use the time-evolved observables
for a finite but nonzero time to complete the argument for justification.
%In order for this to be allowed,
This still requires that
the probability distribution of the drift term
should be suppressed, at least, exponentially at large magnitude.
%Thus we arrive at a simple criterion 

We also point out that 
the integration by parts, which was considered to be the main issue
in justifying the CLM, requires a slightly weaker condition 
than the one we obtain above.
This conclusion is reached by reformulating
the argument starting with
a discretized Langevin time\footnote{Some preliminary discussions 
for finite $\epsilon$
are given already in our previous publication \cite{Nagata:2015uga}
for the purpose of treating the case in which the gauge cooling
transformation remains finite in the $\epsilon\rightarrow 0$ limit.
%However, the subtlety of the $\epsilon\rightarrow 0$ limit
%has not been recognized there.
}
with the step-size $\epsilon$.
In this case, we can always
define the time-evolution of observables
in such a way that it is equivalent to the usual description
with fixed observables and the time-dependent
probability distribution of the complexified variables.
However, an issue arises when one tries to take the 
$\epsilon\rightarrow 0$ limit.
Thus the failure of the integration by parts can be understood
as the failure of the $\epsilon\rightarrow 0$ limit for an 
expression involving the time-evolved observables.
Based on this understanding, we find that
the integration by parts can be justified
if the probability distribution of the drift term
falls of faster than any power-law at large magnitude.
This is slightly weaker than
the condition 
that the probability distribution of the drift term
should be suppressed, at least, exponentially at large magnitude.
%for the validity of using time-evolved observables
%for a finite time.
%% Thus the condition
%% for the validity of the integration by parts
%% is slightly weaker than
%% the condition for the validity of using time-evolved observables
%% for a finite time.
%Therefore, the above condition
%is implies that the latter 
Therefore, we may regard the latter as
%which therefore gives
a necessary and sufficient condition for justifying the CLM.
%Therefore, the condition that the probability distribution of the drift term
%should be suppressed exponentially at large magnitude
%is actually 
In the case of the real Langevin method \cite{Parisi:1980ys},
there is no need to consider the time-evolved observables in 
justifying the method, which implies that all the conditions 
encountered above are simply irrelevant.
%taking into account the subtlety of the limit.

%% In that case, it is not possible to complete the argument
%% for justification, and the CLM loses its basis.
%% On the other hand, 
%% when the probability for the appearance of a large drift term 
%% is suppressed exponentially, the $\epsilon$-expansion is valid
%% and the argument for justification goes through.
%%
%% Thus we arrive at 
%% the necessary and sufficient condition
%% a very simple criterion for justification
%% of the CLM.

%% In the previous argument with the continuous Langevin time,
%% the problem appeared from
%% the integration by parts used in 
%% translating the time evolution of the probability distribution
%% of the complexified variables into that of the observables.
%% In our argument, however, 
%% the translation of the time evolution is possible
%% as long as the expectation values at each Langevin time
%% is well-defined.
%% Therefore, this part of the argument does not give rise to
%% additional conditions for justification.

We substantiate our argument by investigating two simple examples.
%,in which the CLM was considered to fail for different reasons.
The first one is a model studied in ref.~\cite{Nishimura:2015pba}
to clarify the problem related to a singular drift, while
the second one is a model studied in ref.~\cite{Aarts:2013uza}
to clarify the problem related to long excursions 
into the deeply imaginary regime.
In both models, there are two parameter regions;
the CLM works in one of them but fails in the other.
We measure the probability distribution 
%for the magnitude 
of the drift term
and investigate its asymptotic behavior at large magnitude.
It is found that the probability distribution
is indeed exponentially suppressed when the CLM works, 
while it is only power-law suppressed when the CLM fails.
Thus, our simple condition
%provides the necessary and sufficient condition 
%for the method to work,
tells us
clearly
%which can be used to judge 
whether the results obtained by the method are trustable or not
in a unified manner.

The rest of this paper is organized as follows.
In section \ref{sec:0d-model}
we discuss the justification of the CLM,
%% starting with a finite step-size $\epsilon$ for the Langevin time,
%% and show
%% that the $\epsilon \rightarrow 0$ limit is actually subtle.
and point out that the use of time-evolved observables can be subtle.
This leads to our proposal of a necessary and sufficient
condition for justifying the CLM.
In section \ref{sec:examples}
we investigate two models, in which the CLM was thought to fail 
for different reasons. In particular, we show that our new condition can
tell whether the results are trustable or not.
In section \ref{sec:lattice}
we extend the argument in section \ref{sec:0d-model}
to the case of lattice gauge theory.
We also discuss a new possibility for the gauge cooling, which can
reduce the magnitude of the drift term directly.
Section \ref{sec:conclusion}
is devoted to a summary and discussions.
%in a straightforward manner.

%%%%%%%%%%%%%%%%%%%%%%%%%%%%%%%%%%%%%%%%%%%%%%%%%%%%%%%%%%%%%%%%%%%%%%
%%%%%%%%%%%%%%%%%%%%%%%%%%%%%%%%%%%%%%%%%%%%%%%%%%%%%%%%%%%%%%%%%%%%%%
%%%%%%%%%%%%%%%%%%%%%%%%%%%%%%%%%%%%%%%%%%%%%%%%%%%%%%%%%%%%%%%%%%%%%%\
\section{The case of a 0-dimensional model}
\label{sec:0d-model}
%\hspace{0.51cm}
%%%%%%%%%%%%%%%%%%%%%%%%%%%%%%%%%%%%%%%%%%%%%%%%%%%%%%%%%%%%%%%%%%%%%%
%%%%%%%%%%%%%%%%%%%%%%%%%%%%%%%%%%%%%%%%%%%%%%%%%%%%%%%%%%%%%%%%%%%%%%
%%%%%%%%%%%%%%%%%%%%%%%%%%%%%%%%%%%%%%%%%%%%%%%%%%%%%%%%%%%%%%%%%%%%%%

In this section we revisit the argument for justification of the CLM.
In particular, we point out that 
the use of time-evolved observables, which play a crucial role
in the argument, can be subtle, and this
leads to a condition that the probability distribution of
the drift term should fall off exponentially or faster
at large magnitude.
Our argument starts with a finite step-size $\epsilon$ 
for the discretized Langevin time,
which is different from
the previous argument \cite{Aarts:2009uq,Aarts:2011ax},
which starts from the complex Langevin equation 
with a continuous Langevin time.
The purpose of this is to clarify the condition for the validity
of the integration by parts, which was considered the main issue
in the previous argument.
%We find that the $\epsilon \rightarrow 0$ limit 
In fact, we find that this condition is slightly weaker than 
the one we newly obtain.
Therefore, the latter is actually 
a necessary and sufficient condition for the CLM to be justified.
%for justifying the use of time-evolved observables for a finite time.
Here we discuss a 0-dimensional model for simplicity,
%which is much simpler than 
but generalization to the lattice gauge theory
is straightforward as we
%explained in the Appendix \ref{sec:lattice}.
show explicitly in section \ref{sec:lattice}.

We include the gauge cooling procedure
to keep our discussion as general as possible.
This part is similar to what we have already done in our
previous paper \cite{Nagata:2015uga}.
The readers who are not interested in the gauge cooling
can omit the gauge cooling procedure by simply
setting the transformation matrix $g$ to identity in all 
the expressions below.
In ref.~\cite{Nagata:2015uga}, we have also reviewed
the previous argument for justification of the CLM,
which may be compared with our new argument.

%% As we discussed in ref.~\cite{Nagata:2015uga},
%% the gauge cooling can be useful also in a non-gauge system 
%% with a continuous symmetry with the degrees of freedom
%% which has the same order of magnitude as that of the dynamical variables.

\subsection{The complex Langevin method}
\label{sec:CLM}

%\hspace{0.51cm}
%%%%%%%%%%%%%%%%%%%%%%%%%%%%%%%%%%%%%%%%%%%%%%%%%%%%%%%%%%%%%%%%%%%%%%

Let us consider a system of $N$ real variables $x_k$ 
($k=1,\cdots ,N$) given by the partition function\footnote{In many examples, 
the weight is given by $w(x)=\ee^{-S(x)}$ 
in terms of the action $S(x)$, but we prefer not to use the action
in our discussion to avoid any ambiguities arising from
taking the log of 
the complex weight \cite{Mollgaard:2013qra,Mollgaard:2014mga,Greensite:2014cxa}.}
\begin{alignat}{3}
Z = \int dx \, w(x) =  \int \prod_{k} dx_k \, w(x) \ , 
  \label{eq:part-fn}
\end{alignat}
where the weight $w(x)$ is a complex-valued function 
of the real variables $x_k$ ($k=1,\cdots ,N$).

When one considers the Langevin equation for this system, the drift term
\begin{alignat}{3}
%v_k(x) &= -  \frac{\del S(x)}{\del x_k} \ .
v_k(x) &= \frac{1}{w(x)} \frac{\del w(x)}{\del x_k}
\label{eq:def-drift-term}
\end{alignat}
becomes complex, and therefore, one necessarily has to
complexify the dynamical variables\footnote{In this 
respect, there is a closely related approach based on
the so-called Lefschetz 
thimble \cite{Witten:2010cx,Cristoforetti:2012su},
which has attracted much attention recently.
See refs.\cite{Cristoforetti:2013wha,%
Fujii:2013sra,Mukherjee:2014hsa,DiRenzo:2015foa,Fukushima:2015qza} 
and references therein.
There is also a new proposal \cite{Alexandru:2015sua}
for generalizing this approach
to overcome a few important 
problems in the original idea.}
as $x_k  \mapsto z_k = x_k + i y_k$.
Then, the discretized complex Langevin equation is given by
\begin{alignat}{3}
z_k^{(\eta)} (t+\epsilon) 
= z_k^{(\eta)} (t) 
+ \epsilon \, v_k (z)
%- \epsilon \frac{\del S}{\del z_k} 
+ \sqrt{\epsilon} \, \eta_k(t)  \ ,
\label{eq:Langevin-discretized2-complexified}
\end{alignat}
where the drift term $v_k (z)$ is obtained 
by analytically continuing (\ref{eq:def-drift-term}).
The probabilistic variables $\eta_k(t)$ in 
%(\ref{eq:Langevin-discretized2-complexified-cooled})
(\ref{eq:Langevin-discretized2-complexified})
are, in general, complex
\begin{alignat}{3}
\eta_k(t)=\eta^{({\rm R})}_k(t)+ i \eta^{({\rm I})}_k(t) \ ,
\label{eq:complex-noise}
\end{alignat}
and obey the probability distribution
$\propto \ee^{-\frac{1}{4} \sum_t \, 
\{ \frac{1}{N_{\rm R}}\eta_k^{({\rm R})}(t)^2
+\frac{1}{N_{\rm I}}\eta_k^{({\rm I})}(t)^2 \} } $, where
%obey the probability distribution
we have to choose
\begin{alignat}{3}
N_{\rm R} -N_{\rm I} = 1  \ .
\label{NR-NI}
\end{alignat}
For practical purposes, one should actually use
$N_{\rm R}=1$, $N_{\rm I} = 0$,
corresponding to real $\eta_k (t)$,
to reduce the excursions
in the imaginary directions, which
spoil the validity of the method \cite{Aarts:2009uq,Aarts:2011ax,Aarts:2013uza}.

Let us define
the expectation value $\langle \ \cdots \ \rangle_{\eta}$
with respect to $\eta$ as
\begin{alignat}{3}
\langle \ \cdots \ \rangle_{\eta}
= \frac{\int {\cal D}\eta \cdots 
\ee^{
-\frac{1}{4} \sum_t \, 
\{ \frac{1}{N_{\rm R}}\eta_k^{({\rm R})}(t)^2
+\frac{1}{N_{\rm I}}\eta_k^{({\rm I})}(t)^2 \}
}
}
{\int {\cal D}\eta  \, \ee^{-\frac{1}{4} \sum_t \, 
\{ \frac{1}{N_{\rm R}}\eta_k^{({\rm R})}(t)^2
+\frac{1}{N_{\rm I}}\eta_k^{({\rm I})}(t)^2 \}
%\eta(t)^2
}} 
\ .
\label{def-EV-eta-complex}
\end{alignat}
With this notation, we have, for instance,
\begin{alignat}{3}
\Big\langle \eta^{({\rm R})}_k(t_1) \, \eta^{({\rm R})}_l (t_2)  
\Big\rangle_{\eta}
&= 2 N_{\rm R} \, \delta_{kl} \, \delta_{t_1 ,t_2} \ , \nonumber \\
\Big\langle \eta^{({\rm I})}_k(t_1) \, \eta^{({\rm I})}_l (t_2)  
\Big\rangle_{\eta}
&= 2 N_{\rm I} \, \delta_{kl} \, \delta _{t_1  ,t_2} \ , \nonumber \\
\Big\langle \eta_k^{({\rm R})}(t_1) \, \eta^{({\rm I})}_l (t_2)  
\Big\rangle_{\eta}
&= 0 \ .
\label{2pt-corr-eta-complex}
\end{alignat}

When the system (\ref{eq:part-fn}) has a symmetry under
\begin{alignat}{3}
x_j ' = g_{jk} x_k \ ,
\label{symmetry}
\end{alignat}
where $g$ is a representation matrix of a Lie group,
we can use the symmetry to apply gauge cooling.
Upon complexifying the variables $x_k  \mapsto z_k$,
the symmetry property of the drift term and the observables 
naturally enhances from (\ref{symmetry}) to
\begin{alignat}{3}
z_j ' = g_{jk} z_k \ ,
\label{symmetry-complexified}
\end{alignat}
where $g$ is an element of the Lie group that can be obtained
by complexifying the original Lie group.
The discretized complex Langevin equation including
the gauge cooling is given by
\begin{alignat}{3}
\tilde{z}_k^{(\eta)} (t) &= 
g_{kl}
%(z_l^{(\eta)} (t))
%(\tilde{z}^{(\eta)} (t+\epsilon),\tilde{z}^{*(\eta)} (t+\epsilon))
\, z_l^{(\eta)} (t) \ ,  
\label{eq:Langevin-discretized2-complexified-cooled0} \\
%\nonumber \\
z_k^{(\eta)} (t+\epsilon) 
&=   \tilde{z}_k^{(\eta)} (t) 
+ \epsilon \, v_k(\tilde{z}^{(\eta)} (t) )
+ \sqrt{\epsilon} \, \eta_k(t)  \ .
\label{eq:Langevin-discretized2-complexified-cooled}
\end{alignat}
Eq.~(\ref{eq:Langevin-discretized2-complexified-cooled0})
represents the gauge cooling,
where $g$ is an element of the complexified Lie group
chosen appropriately as a function of the configuration 
$z^{(\eta)}(t)$ before cooling.
We regard 
(\ref{eq:Langevin-discretized2-complexified-cooled0}) and
(\ref{eq:Langevin-discretized2-complexified-cooled})
as describing
the $t$-evolution of $z_k^{(\eta)} (t)$ and treat
$\tilde{z}_k^{(\eta)} (t)$ as an intermediate object.
The basic idea is to determine $g$
in such a way that the modified Langevin process 
does not suffer from the problem of the original 
Langevin process (\ref{eq:Langevin-discretized2-complexified}).

We consider observables ${\cal O}(x)$,
which are invariant under (\ref{symmetry})
and admit holomorphic extension to ${\cal O}(x+iy)$.
Note that the symmetry of the observables also
enhances to (\ref{symmetry-complexified}).
Its expectation value can be defined as
\begin{alignat}{3}
\Phi(t) = 
\Big\langle  {\cal O}\Big(x^{(\eta)} (t)+i y^{(\eta)} (t)\Big)
\Big\rangle_{\eta} 
=
\int 
dx \, dy \, {\cal O}(x+iy) \, P(x,y;t) \ ,
\label{OP-rewriting}
\end{alignat}
where we have defined
the probability distribution of $x_k^{(\eta)} (t)$
and $y_k^{(\eta)} (t)$ by
\begin{alignat}{3}
P(x,y;t) = \Bigl\langle \prod_k \delta \Big(x_k - x_k^{(\eta)} (t) \Big)
\, \delta \Big(y_k - y_k^{(\eta)} (t) \Big)
\Bigr \rangle_\eta \ .
\label{def-P-xy}
\end{alignat}
%The justification of the CLM implies that the equality
Under certain conditions, we can show that
\begin{alignat}{3}
%\lim_{\epsilon \rightarrow 0 }
\lim_{t \rightarrow \infty} 
\lim_{\epsilon \rightarrow 0 } \, 
\Phi(t)
&= 
\frac{1}{Z} \int 
dx
%\prod_k dx_k 
\, {\cal O}(x) \, w(x) \ ,
\label{O-time-av-complex}
\end{alignat}
which implies that the CLM is justified.
%, which implies that the CLM 
%
%in the $\epsilon \rightarrow 0$ limit.
%
%Below we discuss the necessary and sufficient condition
%for this statement.
%
%Then one can show under certain conditions that

\subsection{The $t$-evolution of the expectation value}
\label{sec:t-evolution}

Let us first discuss the $t$-evolution of 
the expectation value $\Phi(t)$, which is given by
%Let us therefore consider
\begin{alignat}{3}
\Phi(t + \epsilon) = 
\Big\langle  {\cal O}\Big(x^{(\eta)} (t+\epsilon)+i y^{(\eta)} (t+\epsilon) \Big)
\Big\rangle_{\eta} 
=
\int 
dx \, dy \, {\cal O}(x+iy) \, P(x,y;t+\epsilon) \ .
\label{OP-rewriting-P}
\end{alignat}
Note that the $t$-evolution of $P(x,y;t)$ can be readily obtained from 
the complex Langevin equation 
(\ref{eq:Langevin-discretized2-complexified-cooled0}) and
(\ref{eq:Langevin-discretized2-complexified-cooled}) as
\begin{alignat}{3}
P(x,y;t+\epsilon)
=& \frac{1}{{\cal N}}\int d\eta \, 
\ee^{-\frac{1}{4}  \, 
\{ \frac{1}{N_{\rm R}} \eta_k^{({\rm R})2}
+\frac{1}{N_{\rm I}}\eta_k^{({\rm I})2} \} } \int d\tilde{x} d\tilde{y}
\nonumber \\
& \times 
\delta\Big(x-\tilde{x}
-\epsilon{\rm Re}v(\tilde{z})-\sqrt{\epsilon}\eta^{({\rm R})} \Big) 
\delta\Big(y-\tilde{y}
-\epsilon{\rm Im}v(\tilde{z})-\sqrt{\epsilon}\eta^{({\rm I})} \Big)
\tilde{P}(\tilde{x},\tilde{y};t) 
\nonumber \\
=& \frac{1}{\epsilon{\cal N}} \int d\tilde{x} d\tilde{y}
\exp \left[
-\left\{ 
\frac{\Big(x-\tilde{x}-\epsilon{\rm Re}v(\tilde{z})\Big)^2}{4 \epsilon N_{\rm R}}
+
\frac{\Big(y-\tilde{y}-\epsilon{\rm Im}v(\tilde{z})\Big)^2}{4 \epsilon N_{\rm I}}
\right\}
\right] 
\nonumber \\
& \quad \times \tilde{P}(\tilde{x},\tilde{y};t) \ ,
\label{P-evolve}
\end{alignat}
where ${\cal N}=2 \pi  \sqrt{N_{\rm R} N_{\rm I}} $
is just a normalization constant, and
we have defined the probability distribution for 
$\tilde{z}^{(\eta)}(t)$ in
(\ref{eq:Langevin-discretized2-complexified-cooled0}) as
\begin{alignat}{3}
\tilde{P}(\tilde{x},\tilde{y};t)
&= \int dx dy \,
%d\eta \, 
\delta\Big(\tilde{x}-{\rm Re}(z^{(g)})\Big)
\delta\Big(\tilde{y}-{\rm Im}(z^{(g)})\Big)
P(x,y;t)  \ ,
\label{tilde-P}
\\
z_k^{(g)} & = g_{kl}(x,y) \, z_l \ .
\end{alignat}
%using we have introduced $z_k^{(g)} = g_{kl}(x,y) \, z_l$.
%%
%% We assume that this is a well-defined quantity\footnote{This assumption 
%% is violated, for instance, when $v(z)$ has a singularity at
%% $z=x_{*}+i y_{*}$,
%% and $P(x,y;t)$ is non-zero at $(x,y)=(x_{*} , y_{*})$.
%% In that case, the CLM would fail more miserably because one cannot obtain
%% a finite result. In this paper, we are concerned with a situation
%% in which one obtains a finite result, but it is wrong in the sense
%% that (\ref{O-time-av-complex}) does not hold.}.
Using (\ref{P-evolve}) in (\ref{OP-rewriting-P}),
we obtain
\begin{alignat}{3}
\Phi(t + \epsilon)  &= 
\int 
dx \, dy \, {\cal O}(x+iy) 
 \int d\tilde{x} d\tilde{y} \, \tilde{P}(\tilde{x},\tilde{y};t) 
\nonumber \\
& \quad \times 
 \frac{1}{\epsilon {\cal N}}
\exp \left[
-\left\{ 
\frac{\Big(x-\tilde{x}-\epsilon{\rm Re}v(\tilde{z})\Big)^2}{4 \epsilon N_{\rm R}}
+
\frac{\Big(y-\tilde{y}-\epsilon{\rm Im}v(\tilde{z})\Big)^2}{4 \epsilon N_{\rm I}}
\right\}
\right] \ .
\label{OP-rewriting-P2prev}
\end{alignat}

Here we make an important assumption.
Let us note that
the convergence of the integral (\ref{OP-rewriting})
or (\ref{OP-rewriting-P2prev})
is not guaranteed because 
the observable $|{\cal O}(x+iy)|$ can become infinitely large,
and therefore it is possible that
the expectation value of ${\cal O}(x+iy)$ is ill-defined.
We restrict the observables to those
for which the integral (\ref{OP-rewriting}) converges absolutely 
at any $t\ge 0$.
%In other words, we assume that $P(x,y;t)$ is suppressed strongly
%at $(x,y)$ which makes $|{\cal O}(x+iy)|$ large.
This is legitimate since we are concerned with a situation
in which one obtains a finite result, but it is wrong in the sense
that (\ref{O-time-av-complex}) does not hold.
%it is legitimate to restrict ourselves to the
%cases in which this assumption is satisfied.
%
%these two assumptions are violated,
%the CLM would fail more miserably because one cannot obtain
%a finite result. 

%the CLM 
Under the above assumption, we can exchange the order of integration 
in (\ref{OP-rewriting-P2prev}) due to Fubini's theorem, and rewrite it as
\begin{alignat}{3}
\Phi(t + \epsilon)  &= 
\int 
dx \, dy \, {\cal O}_{\epsilon}(x+iy) \, 
\tilde{P}(x,y;t) \ ,
\label{OP-rewriting-P2}
\end{alignat}
where we have defined
\begin{alignat}{3}
{\cal O}_{\epsilon}(z) 
&= 
 \frac{1}{\epsilon {\cal N}}
\int d\tilde{x} d\tilde{y} \, 
\exp \left[
-  \, 
\left\{  
\frac{\Big(\tilde{x}-x-\epsilon{\rm Re}v(z)\Big)^2}{4\epsilon N_{\rm R} }
+
\frac{\Big(\tilde{y}-y-\epsilon{\rm Im}v(z)\Big)^2}{4\epsilon N_{\rm I}}
\right\}
\right]
{\cal O}(\tilde{x}+i\tilde{y})  
% \ .
\nonumber \\
&=   \frac{1}{ {\cal N}}
\int d\eta \, 
\ee^{-\frac{1}{4}  \, 
\{ \frac{1}{N_{\rm R}} \eta_k^{({\rm R})2}
+\frac{1}{N_{\rm I}}\eta_k^{({\rm I})2} \} } 
O\Big(z+\epsilon \, v(z)+\sqrt{\epsilon}\, \eta  \Big) \ .
\label{OP-rewriting-P3}
\end{alignat}
Note that if ${\cal O}(z)$ and $v_k(z)$ are holomorphic,
so is ${\cal O}_{\epsilon}(z)$. When we say ``holomorphic'',
we admit the case in which the function has singular points.
%% Since these points have measure zero in the integral,
%% they do not make the integral ill-defined.
%%
%% Let us assume that we can expand the expression
%% (\ref{OP-rewriting-P2}) 
%% with respect to $\epsilon$.

In order to proceed further, we expand
(\ref{OP-rewriting-P3}) with respect to $\epsilon$ 
and perform the integration over $\eta$.
After some algebra, we get 
(See Appendix A of ref.~\cite{Nagata:2015uga} for derivation)
\begin{alignat}{3}
{\cal O}_{\epsilon}(z)
&= 
\mbox{\bf :} \ee^{\epsilon L} \mbox{\bf :} \, {\cal O}(z)    \ ,
\label{O-t-evolve-expand}
\end{alignat}
where the expression $\ee^{\epsilon L}$ is a short-hand notation for
\begin{alignat}{3}
\ee^{\epsilon L}
\equiv \sum_{n=0}^{\infty}
  \frac{1}{n!} \, \epsilon^n  L^n \ ,
\label{exp-L}
\end{alignat}
and the operator $L$ is defined by
\begin{alignat}{3}
L &= \left(
%- {\rm Re} \left(\frac{\del S}{\del z_k}\right)
{\rm Re} \, v_k (z)
+ N_{\rm R} \frac{\del}{\del x_k}
\right)
\frac{\del}{\del x_k}
+ \left(
%-   {\rm Im} \left(\frac{\del S}{\del z_k}\right) 
{\rm Im}  \, v_k (z)
+ N_{\rm I} \frac{\del}{\del y_k}
\right)
\frac{\del}{\del y_k} \ .
\label{L-expression}
\end{alignat}
The symbol $\mbox{\bf :} \ldots \mbox{\bf :}$ in (\ref{O-t-evolve-expand})
implies that the operators are ordered in such a way that
derivative operators appear on the right; e.g.,
$\mbox{\bf :} ( f(x) + \del)^2 \mbox{\bf :}= f(x)^2 + 2f(x)\del + \del^2$.

Since ${\cal O}(z)$ is a holomorphic function of $z$, we have
\begin{alignat}{3}
%% \frac{\del f}{\del x_k} &= \frac{\del f}{\del y_k}
%% \frac{\del f}{\del y_k} &= \frac{\del f}{\del x_k}
L {\cal O}(z) &= \left(
%- {\rm Re} \left(\frac{\del S}{\del z_k}\right)
{\rm Re} \, v_k (z)
+ N_{\rm R} \frac{\del}{\del z_k} 
\right)
\frac{\del {\cal O}}{\del z_k}
+ \left(
% -  {\rm Im} \left(\frac{\del S}{\del z_k}\right) 
{\rm Im} \,  v_k (z)
+ i N_{\rm I} \frac{\del}{\del z_k}
\right)
\left( i \frac{\del {\cal O}}{\del z_k} \right)  \nonumber \\
&= \left(
%-  \frac{\del S}{\del z_k}
 v_k (z) 
+ ( N_{\rm R} - N_{\rm I})  \frac{\del}{\del z_k} 
\right)
\frac{\del {\cal O}}{\del z_k}
\nonumber \\
 &= \tilde{L} {\cal O}(z) \ ,
\label{LO}
\end{alignat}
where we have used (\ref{NR-NI}) and defined
\begin{alignat}{3}
\tilde{L} 
%  &= \left( \frac{\del}{\del z_k} - \frac{\del S}{\del z_k} 
&= \left( \frac{\del}{\del z_k} + v_k(z)
\right)
\frac{\del }{\del z_k} \ .
\label{L-tilde}
\end{alignat}
Hence we can rewrite (\ref{O-t-evolve-expand}) as
\begin{alignat}{3}
{\cal O}_{\epsilon}(z)
&= \mbox{\bf :} \ee^{\epsilon \tilde{L}} \mbox{\bf :} \, {\cal O}(z)    \ .
\label{O-t-evolve-expand2}
\end{alignat}
%where $t = n \epsilon$.

Plugging (\ref{O-t-evolve-expand2}) in (\ref{OP-rewriting-P2}),
we formally obtain
\begin{alignat}{3}
\Phi(t + \epsilon)  &= 
\sum_{n=0}^{\infty}
  \frac{1}{n!} \, \epsilon^n  
\int 
dx \, dy \, 
\Big( \mbox{\bf :} \tilde{L}^n  \mbox{\bf :} \,  {\cal O}(z) \Big)
\tilde{P}(x,y;t) 
\nonumber \\
&= 
\sum_{n=0}^{\infty}
  \frac{1}{n!} \, \epsilon^n  
\int 
dx \, dy \, 
\left.
\Big( \mbox{\bf :} \tilde{L}^n  \mbox{\bf :} \,  {\cal O}(z) \Big) 
\right|_{z^{(g)}}
P(x,y;t) 
\nonumber \\
&= 
\sum_{n=0}^{\infty}
  \frac{1}{n!} \, \epsilon^n  
\int 
dx \, dy \, 
\Big( \mbox{\bf :} \tilde{L}^n  \mbox{\bf :} \,  {\cal O}(z) \Big) 
P(x,y;t) \ .
\label{OP-rewriting-P3b}
\end{alignat}
In the third equality, we have used the fact that 
$\mbox{\bf :} \tilde{L}^n  \mbox{\bf :} \,  {\cal O}(z)$ are 
invariant under the 
%O($N,\bbC$) 
complexified symmetry
transformation (\ref{symmetry-complexified}).
%% From (\ref{OP-rewriting-P3b})
%% it follows that $\lim_{\epsilon \rightarrow 0} \Phi(t)$ is
%% differentiable, and its derivative is given by
%% \begin{alignat}{3}
%% \frac{d}{dt} \, \lim_{\epsilon \rightarrow 0}  \Phi(t) 
%% &=  \lim_{\epsilon \rightarrow 0}  \int 
%% dx \, dy \, 
%% \Big\{ \tilde{L} \, {\cal O}(z) \Big\} \, P(x,y;t)  \ .
%% \label{OP-rewriting-P3c}
%% \end{alignat}
Thus we find \cite{Nagata:2015uga}
that the effect of the gauge cooling represented by $g$
disappears in the $t$-evolution of 
%the O($N,\bbC$) invariant observables,
%observables invariant under the symmetry transformation (\ref{symmetry}),
observables invariant under the symmetry transformation (\ref{symmetry-complexified}),
although the $t$-evolution of the probability distribution $P(x,y;t)$
is affected nontrivially by the gauge cooling as in (\ref{P-evolve}).

If the $\epsilon$-expansion (\ref{OP-rewriting-P3b})
is valid, we can truncate the infinite series 
for sufficiently small $\epsilon$ as
\begin{alignat}{3}
\Phi(t + \epsilon)
&= \Phi(t) + \epsilon \int 
dx \, dy \, 
\Big\{ \tilde{L} \, {\cal O}(z) \Big\}
\, P(x,y;t) + O(\epsilon^2) \ ,
\label{OP-rewriting-P3b-truncate}
\end{alignat}
which implies that the $\epsilon\rightarrow 0$ limit
can be taken without any problem, and we get
\begin{alignat}{3}
\frac{d}{dt} \, \Phi(t)
&=  \int dx \, dy \, 
\Big\{ \tilde{L} \, {\cal O}(z) \Big\}
\, P(x,y;t)  \ .
\label{OP-rewriting-P3b-cont-lim}
\end{alignat}
%The rest of the argument is the same 
%as in refs.~\cite{Aarts:2009uq,Aarts:2011ax}.

However, it is known 
from the previous argument \cite{Aarts:2009uq,Aarts:2011ax}
using a continuous Langevin time
that there are cases
in which (\ref{OP-rewriting-P3b-cont-lim}) does not 
hold due to the failure of the integration by parts.
In the present argument, 
the reason why (\ref{OP-rewriting-P3b-cont-lim}) can be violated
should be attributed to the possible breakdown of 
the expression (\ref{OP-rewriting-P3b}).
Note that the operator $\tilde{L}^n$ involves the $n$th power of
the drift term $v_k(z)$ in (\ref{L-tilde}),
which may become infinitely large.
Therefore, the integral that appears in (\ref{OP-rewriting-P3b})
may be divergent for large enough $n$.

We emphasize here that what we have done in this section is just
an alternative presentation of the known problem that
(\ref{OP-rewriting-P3b-cont-lim}) can be violated.
In particular, 
the previous argument using a continuous Langevin time 
is absolutely correct since
the discretized complex Langevin equation
approaches smoothly the continuum one 
in the $\epsilon\rightarrow 0$ limit.
%Nevertheless, the expression (\ref{OP-rewriting-P3b})
%can be invalid when the integral 
Note also that the problem under discussion 
cannot be solved by using a sufficiently small $\epsilon$ or 
an adaptive step-size \cite{Aarts:2009dg}.
The advantage of our argument using a discretized Langevin time 
is that we can interpret the failure of the integration by parts
in the previous argument as the breakdown of 
the $\epsilon$-expansion (\ref{OP-rewriting-P3b}) due to the 
appearance of a large drift term. This makes it possible to 
compare the condition required for the validity of the
expression (\ref{OP-rewriting-P3b-cont-lim})
with the one discussed in the next section.

%\subsection{Proof of the key identity (\ref{O-time-av-complex})}
\subsection{Subtlety in the use of time-evolved observables}
\label{sec:key-id}

In this section 
we assume that
the problem discussed in the previous section
does not occur and that (\ref{OP-rewriting-P3b-cont-lim}) holds.
Repeating this argument for $\tilde{L}^n \, {\cal O}(z)$, we 
obtain
\begin{alignat}{3}
\left( \frac{d}{dt} \right)^n \, \Phi(t)
&=  \int dx \, dy \, 
\Big\{ \tilde{L}^n \, {\cal O}(z) \Big\}
\, P(x,y;t)  \ .
\label{OP-rewriting-P3b-cont-lim-Ln}
\end{alignat}
Therefore, a finite time-evolution can be written
formally as\footnote{Subtlety of 
eq.~(\ref{OP-rewriting-P3b-cont-lim-exp}) for finite $\tau$ at $t=0$
was discussed in ref.~\cite{Duncan:2012tc}
in a one-variable case with a complex quartic action.
We thank M.~Niedermaier for bringing our attention to this work.}
\begin{alignat}{3}
 \Phi(t+\tau)
&=  \sum_{n=0}^{\infty}
  \frac{1}{n!} \, \tau^n
\int dx \, dy \, 
\Big\{  \tilde{L}^n \, {\cal O}(z) \Big\}
\, P(x,y;t)  \ ,
%%  \Phi(t+\delta t) 
%% &=  \int dx \, dy \, 
%% \Big\{ \ee^{\delta t \, \tilde{L}} \, {\cal O}(z) \Big\}
%% \, P(x,y;t)  \ ,
\label{OP-rewriting-P3b-cont-lim-exp}
\end{alignat}
%In order for this expression to make 
which is similar to (\ref{OP-rewriting-P3b}).
%and the validity of the $\tau$-expansion is not guaranteed.
In order for this expression to be valid for a finite $\tau$, however,
it is not sufficient to assume that
the integral that appears in (\ref{OP-rewriting-P3b-cont-lim-exp})
is convergent for arbitrary $n$.
What matters is 
%It depends on 
the convergence radius of the 
infinite series (\ref{OP-rewriting-P3b-cont-lim-exp}).
%obtained after the integral
In the previous argument, 
the proof of the key identity (\ref{O-time-av-complex})
was given assuming implicitly that the convergence radius is infinite.
This is actually a too strong assumption, which is not satisfied
even in cases where the CLM is known to give correct 
results (See, e.g., our results in Section \ref{sec:examples}.).
Below we show that we can modify the proof slightly so that
we only have to assume that
the convergence radius $\tau_{\rm conv}(t)$, which depends on $t$ in general,
is bounded from below as $\tau_{\rm conv}(t) \ge \tau_0 > 0$ 
for $0 \le t < \infty$.

In order to show (\ref{O-time-av-complex}), we first 
%consider the following lemma.
%show that 
prove the lemma
%% \footnote{In order to prove (\ref{O-time-av-complex}),
%% refs.~(\ref\cite{Aarts:2009uq,Aarts:2011ax})
%% used the time-evolved operator
%% $ {\cal O}(z;t) = \ee {t \tilde{L}} {\cal O}(z)$,
%% which has to be defined by Taylor expanding the exponential function.
%% However, it is possible that the series we get
%% $ \int dx dy \, 
%% \Big\{ \ee {t \tilde{L}}  {\cal O}(x)  \Big\} \, P(x,y;\tau)
%% = \sum_{k=0}^{\infty}
%%   \frac{1}{k!} \, t^k
%% \int dx dy \, \Big\{ (\tilde{L})^k \, {\cal O}(x+iy)  \Big\} \, P(x,y;\tau) $
%% may not converge at large $t$.
%% }
\begin{alignat}{3}
\int dx dy \, \Big\{ \tilde{L}^n \, {\cal O}(x+iy) \Big\} \, P(x,y;t)
= \int dx \, \Big\{ (L_0)^n \, {\cal O}(x)  \Big\} \, \rho(x;t) 
\label{P-rho-rel}
\end{alignat}
for arbitrary integer $n$ and arbitrary $t\ge 0$,
where the operator $L_0$ is defined by 
\begin{alignat}{3}
%\quad \quad
%\frac{\del}{\del t} {\cal O}(x;t) = L_0 {\cal O}(x;t) \ ,
L_0 &= \left(
\frac{\del}{\del x_k} 
%- \frac{\del S}{\del x_k} 
+v_k(x)
\right)
\frac{\del}{\del x_k} \ ,
\label{L0-expression}
\end{alignat}
and the complex valued function $\rho(x;t)$ is 
defined as the solution to 
the Fokker-Planck (FP) equation
\begin{alignat}{3}
\frac{\del \rho}{\del t}
&= (L_0)^\top  \rho =
\frac{\del}{\del x_k} 
\left( \frac{\del}{\del x_k} -v_k(x)  \right) \rho \ ,
\label{FPeq-complex}
\\
\rho(x;0)& =\rho(x) \ .
\end{alignat}
Here the symbol $L_0^{\top}$ is defined as an operator
satisfying
$\langle L_0 f ,g \rangle=\langle f,L_0^{\top} g \rangle$,
where $\langle f, g \rangle  \equiv 
\int f(x)
g(x) dx$,
assuming that $f$ and $g$ are 
functions that allow integration by parts.
The initial condition is assumed to be
\begin{alignat}{3}
P(x,y;0)=\rho(x) \, \delta(y)  \ ,
\label{P-rho-initial}
\end{alignat}
%$P(x,y;0)=\rho(x,0)\ge 0$
where $\rho(x) \ge 0$ and $\int dx \rho(x) =1 $,
so that (\ref{P-rho-rel}) is trivially satisfied at $t=0$.

The proof of (\ref{P-rho-rel}) is then given by induction 
with respect to $t$.
Let us assume that (\ref{P-rho-rel}) holds at $t=t_0$.
Then we obtain
\begin{alignat}{3}
\int dx dy \, \Big\{ 
\ee^{\tau  \tilde{L}}
 \, {\cal O}(x+iy) \Big\} \, P(x,y;t_0)
&= \int dx \, \Big\{ 
\ee^{\tau  L_0 }
 \, {\cal O}(x)  \Big\} \, \rho(x;t_0) \ ,
\label{etL}
\end{alignat}
where $\tau$ should be smaller than the convergence radius of
the $\tau$-expansion (\ref{OP-rewriting-P3b-cont-lim-exp}) at $t=t_0$.
(The $\tau$-expansion on the right-hand side of (\ref{etL})
is expected to have
no problems due to the properties of
the complex weight $\rho(x;t_0)$ obtained by solving the 
FP equation (\ref{FPeq-complex}) for a well-defined system.)
%assumed to be finite at any $t$.
%
%\footnote{The convergence radius may depend on $t$.}
Since taking the derivative with respect to $\tau$ 
does not alter the convergence radius, we obtain 
\begin{alignat}{3}
\int dx dy \, \Big\{ 
\ee^{\tau  \tilde{L}}
\tilde{L}^n \, {\cal O}(x+iy) \Big\} \, P(x,y;t_0)
&= \int dx \, \Big\{ 
\ee^{\tau  L_0 }
(L_0)^n \, {\cal O}(x)  \Big\} \, \rho(x;t_0)
\label{etL-L0}
\end{alignat}
for arbitrary $n$. Note that
\begin{alignat}{3}
\mbox{l.h.s.\ of eq.~(\ref{etL-L0})} &=
\int dx dy \, \Big\{ 
\tilde{L}^n \, {\cal O}(x+iy) \Big\} \, P(x,y;t_0+\tau) \ ,
\label{P-rho-rel-2}
\end{alignat}
where we have used a relation like
(\ref{OP-rewriting-P3b-cont-lim-exp})
for the observable $\tilde{L}^n \, {\cal O}(x+iy)$, and
\begin{alignat}{3}
\mbox{r.h.s.\ of eq.~(\ref{etL-L0})}
&= \int dx \, \Big\{ 
(L_0)^n \, {\cal O}(x)  \Big\} \,  
\ee^{\tau  (L_0)^\top }\rho(x;t_0) 
\nonumber \\
&= \int dx \, \Big\{ 
(L_0)^n \, {\cal O}(x)  \Big\} \,  
\rho(x;t_0+\tau)  \ ,
\label{P-rho-rel-2.5}
\end{alignat}
where we have used
integration by parts\footnote{This is expected to be valid,
as stated also in refs.~\cite{Aarts:2009uq,Aarts:2011ax},
%because 
%the integration is performed over the original real variables, and 
due to the properties of
the complex weight $\rho(x)$ obtained by solving the 
FP equation (\ref{FPeq-complex}) for a well-defined system.}
in the first equality,
and (\ref{FPeq-complex}) in the second equality. 
Thus we find that (\ref{P-rho-rel}) holds at $t=t_0+\tau$, which
completes the proof of (\ref{P-rho-rel}) for arbitrary $t\ge 0$.

In order to show (\ref{O-time-av-complex}), we only need
to consider the $n=0$ case in (\ref{P-rho-rel}), which reads
\begin{alignat}{3}
\int dx dy \,  {\cal O}(x+iy) \, P(x,y;t)
= \int dx \,  {\cal O}(x)   \, \rho(x;t)  \ .
\label{P-rho-rel-3}
\end{alignat}
Note that eq.~(\ref{FPeq-complex})
has a $t$-independent solution
\begin{alignat}{3}
\rho_{\rm time-indep}(x) = \frac{1}{Z} \, w(x) \ .
\label{time-indep-sol-complex}
\end{alignat}
According to the argument given 
in ref.~\cite{Nishimura:2015pba},
the solution to (\ref{FPeq-complex})
%(\ref{FPeq-complex}) 
asymptotes to
(\ref{time-indep-sol-complex}) at large $t$
if (\ref{P-rho-rel-3}) holds and $P(x,y;t)$ converges
to a unique distribution in the $t\rightarrow \infty$ limit.
Hence, (\ref{O-time-av-complex}) follows from (\ref{P-rho-rel-3}).

%\subsection{Criterion for the validity of the $\epsilon$-expansion}
\subsection{The condition for correct convergence}
\label{sec:criterion}
%\hspace{0.51cm}
%%%%%%%%%%%%%%%%%%%%%%%%%%%%%%%%%%%%%%%%%%%%%%%%%%%%%%%%%%%%%%%%%%%%%%

Let us discuss the condition for the validity of the $\epsilon$-expansion
(\ref{OP-rewriting-P3b})
and the condition for 
the $\tau$-expansion (\ref{OP-rewriting-P3b-cont-lim-exp})
to have a finite convergence radius.
%and the $\tau$-expansion (\ref{OP-rewriting-P3b-cont-lim-exp}).
In fact, it is the latter that is stronger.
% condition because it requires a finite convergence radius.
% and (\ref{OP-rewriting-P-expand}).
As we mentioned in section \ref{sec:t-evolution},
these conditions are
related to the behavior of the probability distribution
for such configurations $(x,y)$ that make the drift term $v_k(z)$ large.
More precisely, we are concerned with the magnitude of the drift term,
which may be defined as
\begin{alignat}{3}
%u(z) =  \max_{\vec{n}} | \vec{n} \cdot \vec{v}(z) | \ ,
u(z) =  \max_{g } \max_{1 \le i \le N} | v_i(z^{(g)}) | \ ,
\label{def-v-magnitude}
\end{alignat}
%where $\vec{v} = (v_1 , \cdots , v_N)$ and $\vec{n}$ is 
%a unit vector $\vec{n}\cdot \vec{n} = 1$.
where $g$ represents a symmetry transformation (\ref{symmetry})
of the original theory.\footnote{In the case of ${\rm O}(N)$ symmetry
$g \in {\rm O}(N)$, for instance, 
the definition (\ref{def-v-magnitude}) is equivalent
to $u(z)= \max_{\vec{n}} |\vec{n} \cdot \vec{v}(z)|$, where
the maximum is taken with respect to a unit vector $\vec{n}$ in $\bbR ^N$.}
Note that $u(z)$ thus defined is invariant under (\ref{symmetry}).
The integral that appears in (\ref{OP-rewriting-P3b})
and (\ref{OP-rewriting-P3b-cont-lim-exp})
% and (\ref{OP-rewriting-P-expand})
for each $n$ involves
\begin{alignat}{3}
\int dx \, dy \, u(z)^n \, P(x,y;t) = \int_0^\infty  du \, u^n \, p(u;t)
\label{simplified-integral}
\end{alignat}
as the most dominant contribution, where we have defined
the probability distribution of the magnitude $u(z)$ by
\begin{alignat}{3}
p(u;t) \equiv  \int dx \, dy \, \delta(u(z)-u) \, P(x,y;t) \ .
\label{def-u-prob}
\end{alignat}
%If the probability distribution $p(u;t)$ is only power-law suppressed 
If $p(u;t)$ is only power-law suppressed 
at large $u$, the integral (\ref{simplified-integral}) is divergent 
for sufficiently large $n$.
Therefore, in order for (\ref{simplified-integral}) 
to be convergent for arbitrary $n$, 
$p(u;t)$ should fall off faster than any power law.
This is required for
% the $\epsilon$-expansion to be valid. 
the $\epsilon$-expansion (\ref{OP-rewriting-P3b})
or the $\tau$-expansion (\ref{OP-rewriting-P3b-cont-lim-exp})
to be valid.

Here we consider the case in which $p(u;t)$ is exponentially suppressed as
$p(u;t)\sim \ee^{-\kappa u}$ at large $u$.
Then, the integral (\ref{simplified-integral}) can be estimated as
\begin{alignat}{3}
\int_0^\infty  du \, u^n \, p(u;t) \sim \frac{n ! }{\kappa^{n+1}} \ .
\label{simplified-integral2}
\end{alignat}
Plugging this into (\ref{OP-rewriting-P3b-cont-lim-exp}),
%(\ref{OP-rewriting-P3b}),
% or (\ref{OP-rewriting-P-expand}),
we find that the convergence radius of the infinite series can be
estimated as $\tau \sim \kappa$.
This implies that $p(u;t)$ has to fall off 
exponentially or faster in order for the convergence radius 
of the $\tau$-expansion (\ref{OP-rewriting-P3b-cont-lim-exp})
to be nonzero,
%which is important in showing eq.~(\ref{P-rho-rel-2}).
%which is important in obtaining eq.~(\ref{etL}).
which is important in our argument given in section \ref{sec:key-id}.

Let us discuss the subtlety of 
the $\epsilon$-expansion (\ref{OP-rewriting-P3b})
in more detail.
Note that $\Phi(t+\epsilon)$ defined by
(\ref{OP-rewriting-P2prev}) is a finite well-defined quantity
for a finite $\epsilon$ under the assumption made below
eq.~(\ref{OP-rewriting-P2prev}).
Nevertheless, the $\epsilon$-expansion
(\ref{OP-rewriting-P3b})
% and (\ref{OP-rewriting-P-expand})
can be ill-defined.
This can happen because the expansion parameter $\epsilon$
is multiplied to the drift term in (\ref{OP-rewriting-P2prev}), 
which can become infinitely large in the integral.
In order to illustrate this point, let us consider a
simple integral
\begin{alignat}{3}
I = \int_{-1}^{1} dx \, \ee^{-\epsilon/x^2}  \ ,
\label{I-def}
\end{alignat}
which is clearly well-defined for arbitrary $\epsilon \ge 0$.
However, if we expand the integrand with respect to $\epsilon$,
we get
\begin{alignat}{3}
I = 
\sum_{n=0}^{\infty}
  \frac{1}{n!} \, \epsilon^n \,  (-1)^n
\int_{-1}^{1} dx \, \frac{1}{x^{2n}} \ ,
\label{I-def-expand}
\end{alignat}
which is invalid because
we obtain divergent terms for $n \ge 1$.
%% In eq.~(\ref{I-def}),
%% the term $1/x^2$ in the exponent diverges at $x=0$,
%% and plays the role of the drift term.
%% In fact, the contribution from $x \ll \sqrt{\epsilon}$ in the integral
%% is strongly suppressed due to the integrand $\ee^{-\epsilon/x^2}$
%% for $\epsilon >0$.

%except for the $O(\epsilon^0)$ term.
%Namely, the $\epsilon$-expansion
We can evaluate (\ref{I-def}) as follows.
Changing the integration variable $t=\sqrt{\epsilon}/x$, we get
\begin{alignat}{3}
I &= 2 \sqrt{\epsilon} 
\int_{\sqrt{\epsilon}}^{\infty} dt \, \frac{1}{t^2} \, \ee^{-t^2} 
%% &= 2 \sqrt{\epsilon} 
%% (\frac{1}{\sqrt{\epsilon}}  \ee^{-\epsilon} 
%% -2 \int_{\sqrt{\epsilon}}^{\infty} dt \,  \ee^{-t^2} )
%% \nonumber \\
= 2  \, 
\{ \ee^{-\epsilon} 
-\sqrt{\pi\epsilon} \, (1 - {\rm Erf}(\sqrt{\epsilon})) \} \ ,
\label{I-def2}
\end{alignat}
where we have performed integration by parts in the first equality,
and ${\rm Erf}$ is the error function.
Expanding (\ref{I-def2}) with respect to $\epsilon$, 
we obtain $O(\epsilon^{n/2})$ terms, which are absent in the
formal expression (\ref{I-def-expand}).

%%%%%%%%%%%%%%%%%%%%%%%%%%%%%%%%%%%%%%%%%%%%%%%%%%%%%%%%%%%%%%%%%%%%%%

%%%%%%%%%%%%%%%%%%%%%%%%%%%%%%%%%%%%%%%%%%%%%%
\subsection{Some comments on the previous argument}
\label{sec:relation-prev-work}
%\hspace{0.51cm}

In this subsection, we clarify the relationship of our new argument and the
previous one. Here we omit the gauge cooling for simplicity.
In ref.~\cite{Aarts:2009uq,Aarts:2011ax},
the quantity
%(\ref{O-time-av-complex}) was proved by 
\begin{alignat}{3}
F(t,\tau) & \equiv
 \int dx dy \, 
{\cal O}(z; \tau) 
%\Big\{ \ee^{\tau \tilde{L}} {\cal O}(z) \Big\}
\, P(x,y;t-\tau) 
\label{def-F}
\end{alignat}
was introduced
with the time-evolved observable
$ {\cal O}(z;\tau) = \ee^{\tau \tilde{L}} {\cal O}(z)$, and it was shown to be
$\tau$-independent for $0 \le \tau \le t$
by using the integration by parts
\begin{alignat}{3}
\int 
dx \, dy \, {\cal O}(z;\tau) \, 
L^{\top}  \, P(x,y;t-\tau)  
%% &= \int dx \, dy \, 
%% \Big\{ \tilde{L} \, {\cal O}(z;\tau) \Big\}
%% \, P(x,y;t-\tau)  
%% \nonumber \\
&=\int dx \, dy \, 
\Big\{ L \, {\cal O}(z;\tau) \Big\}
\, P(x,y;t-\tau)  \ .
\label{nec-suf-condition-old}
\end{alignat}
Note, however, that the quantity (\ref{def-F})
%%where the time-evolved operator
%%$ {\cal O}(z;\tau) = \ee^{\tau \tilde{L}} {\cal O}(z)$ is introduced.
%However, this quantity 
has to be evaluated as
%the time-evolved operator $ {\cal O}(z;t)$
%has to be defined by Taylor expanding the exponential function as
\begin{alignat}{3}
F(t,\tau) 
%% &=
%% \int dx dy \, 
%% \Big\{ \ee ^{\tau \tilde{L}}  {\cal O}(x)  \Big\} \, P(x,y;t-\tau)
&= \sum_{n=0}^{\infty}
  \frac{1}{n!} \, \tau^n
\int dx dy \, \Big\{ \tilde{L}^n \, {\cal O}(z)  \Big\} \, P(x,y;t-\tau)  \ , 
\label{finite-t-evolve}
\end{alignat}
where the infinite series on the right-hand side 
%in the expressions like
may have a finite convergence radius $\tau = \tau_{\rm conv}$.
In that case, (\ref{finite-t-evolve}) is ill-defined for $\tau > \tau_{\rm conv}$.
Our argument in section \ref{sec:key-id}
avoids this problem by using 
%the time-evolved operator $ {\cal O}(z;\tau)$ only for $\tau < \tau_{\rm conv}$,
(\ref{etL}) only for $\tau < \tau_{\rm conv}$
and employing the induction with respect to $t$ instead.
%% $ \ee^{\tau \tilde{L}} {\cal O}(z)$.is defined by
%% a sufficiently small $\tau$
%% in (\ref{etL}).

%% Also we would like to comment on our argument
%% based on the lemma (\ref{P-rho-rel}), which does not appear
%% in the previous argument.

%%%%%%%%
Let us discuss the validity of
the integration by parts (\ref{nec-suf-condition-old}).
Expanding (\ref{P-evolve}) with respect to $\epsilon$, we obtain
\begin{alignat}{3}
P(x,y;t+\epsilon)  
=&  
%(1+  \epsilon L^\top ) \, 
( \mbox{\bf :} \ee^{\epsilon L} \mbox{\bf :}
 )^{\top}  \, 
P(x,y;t) \ .
\label{FPeq-discretized-cmp}
\end{alignat}
%\nonumber \\
In the $\epsilon\rightarrow 0$ limit, we obtain
the FP-like equation
\begin{alignat}{3}
\frac{\del}{\del t} P(x,y;t) &= L^\top P(x,y;t) \ .
\label{FP-like-eq}
\end{alignat}
Using this, we obtain
\begin{alignat}{3}
\frac{\del }{\del t} F(t,\tau)
&=\int dx \, dy \, 
 {\cal O}(z;\tau) \, \frac{\del }{\del t} P(x,y;t-\tau)   \nn \\
&= \int dx \, dy \, 
 {\cal O}(z;\tau) \, L^\top P(x,y;t-\tau)   \ ,
\label{FP-like-eq-in-dF}
\end{alignat}
which is the left-hand side of (\ref{nec-suf-condition-old}).
On the other hand, our argument given before
(\ref{OP-rewriting-P3b-cont-lim}) implies that
\begin{alignat}{3}
\frac{\del }{\del t} F(t,\tau)
&=\int dx \, dy \, 
\Big\{ \tilde{L} \, {\cal O}(z;\tau) \Big\}
\, P(x,y;t-\tau)  
\label{d-dt-F}
\end{alignat}
may or may not hold depending on the validity
of the $\epsilon$-expansion like (\ref{OP-rewriting-P3b}).
Note that the right-hand side of (\ref{d-dt-F})
is nothing but the right-hand side of (\ref{nec-suf-condition-old})
due to (\ref{LO}).
From (\ref{FP-like-eq-in-dF}) and (\ref{d-dt-F}),
we therefore find that
the validity of the integration by parts (\ref{nec-suf-condition-old})
is equivalent to
the validity of (\ref{d-dt-F}),
which requires that
the probability distribution of the drift term
falls of faster than any power-law at large magnitude.
This condition is slightly weaker than the one from the validity
of the use of time-evolved observables for a finite time.
Note that a function $f(x)= e^{-\sqrt{x}}$, for instance,
falls off faster than
any power-law at large $x$, and yet it is not suppressed exponentially
at large $x$.
Therefore, we consider that a necessary and sufficient condition for 
justifying the CLM is that
the probability distribution of the drift term
falls off exponentially or faster at large magnitude.

%% Note, however, that the previous argument used continuous $t$ from the outset,
%% and (\ref{OP-rewriting-P3b-cont-lim-2}) was assumed to hold.
%% According to our new argument starting with discretized $t$,
%% (\ref{OP-rewriting-P3b-cont-lim-2}) does not hold
%% when the $\epsilon$-expansion (\ref{OP-rewriting-P-expand}) is invalid.
%% %Also, the integration by parts in question does not appear in our argument.
%% Therefore, it is misleading to state that
%% the failure of the integration by parts (\ref{nec-suf-condition-old})
%% is the cause of the problem.

In the previous work \cite{Aarts:2009uq,Aarts:2011ax},
it was recognized that the probability distribution of the
complexified dynamical variables should fall off 
fast enough at large absolute values
to make sure that the integration by parts used
in the argument is valid.
However, the rate of the fall-off required to justify 
the CLM was not clear. 
This was also the case with the 
singular-drift problem \cite{Nishimura:2015pba}.
How fast the probability distribution should fall off near
the singularity was not clear.
For this reason,
while it was possible to understand the failure of the CLM
found by comparison with correct results available from other methods,
it was not possible to tell whether the results of the CLM
are trustable or not without knowing the correct results in advance.
The advantage of our condition
based on the probability distribution of the drift term
is that we can clearly state that it is 
the exponential fall-off that is required for justification of the CLM.
This condition ensures 
not only the validity of the integration by parts
used in the argument
but also the validity of the use of time-evolved observables 
for a finite non-zero time.
As we demonstrate in Section \ref{sec:examples},
the condition indeed tells us clearly whether the results of the CLM
are trustable or not.

Let us also comment on the property
\begin{alignat}{3}
\lim_{t\rightarrow \infty}
\int dx \, dy \, 
\Big\{ \tilde{L} \, {\cal O}(z) \Big\}
\, P(x,y;t)  = 0 \ ,
\label{nec-condition-old}
\end{alignat}
which was proposed as a necessary condition 
for justifying the CLM \cite{Aarts:2011ax}.
From the viewpoint of our new argument,
(\ref{nec-condition-old}) follows from
(\ref{OP-rewriting-P3b-cont-lim}),
which is true 
%This follows from (\ref{OP-rewriting-P3b-truncate})
if the $\epsilon$-expansion is valid.
%% Therefore, (\ref{nec-condition-old}) may still be viewed
%% as a necessary condition from the viewpoint of our argument.
However, the quantity on the left-hand side of (\ref{nec-condition-old})
is difficult to evaluate 
since the history of the observable $\tilde{L} \, {\cal O}(z)$ 
typically has spikes with different phase factors,
and huge cancellations occur among configurations.
This limits the usefulness of (\ref{nec-condition-old}) as a
necessary condition.
\subsection{The case of the real Langevin method}
\label{sec:real-langevin}
%\hspace{0.51cm}

In order to appreciate better
the situation in the complex Langevin method,
%let us consider the case of 
let us here consider the case of 
the real Langevin method \cite{Parisi:1980ys},
which is a standard method for a real-action system
based on importance sampling.
In this case, there is no need to complexify the dynamical variables,
and the probability distribution $P(x;t)$ and the weight $\rho(x;t)$
are identical. The discussion in section \ref{sec:key-id} is not 
needed, and therefore 
the expressions like (\ref{OP-rewriting-P3b})
and (\ref{OP-rewriting-P3b-cont-lim-exp})
%(\ref{OP-rewriting-P-expand})
do not have to make sense.
Thus the issues concerning the time-evolved observables become
totally irrelevant.

All we need to justify the method is to show that
the discretized $t$-evolution of $P(x;t)$ like
(\ref{P-evolve}) reduces to the FP equation 
(\ref{FPeq-complex}) in the $\epsilon\rightarrow 0$ limit.
Note that the $\epsilon$-expansion of (\ref{P-evolve})
gives (\ref{FPeq-discretized-cmp}),
and the FP equation is obtained if
the expansion can be truncated at the order of $\epsilon$.
The problem occurs in the region of $x$, where the drift term $v_k(x)$ 
becomes large.
However, the integral of $P(x;t)$ in that region
is typically small, and it is expected to
vanish in the $\epsilon \rightarrow 0$ limit.
Therefore, we may expect that  
(\ref{P-evolve}) reduces to the FP equation 
(\ref{FPeq-complex}) in the $\epsilon\rightarrow 0$ limit.
In order to confirm this, we have studied a system
\begin{eqnarray}
%Z = \int dx \, (x+i\alpha) \, \ee^{-(x-m)^2/2} \ ,
Z = \int dx \, |x|^{-1/2} \, \ee^{-x^2/2} \ ,
%Z = \int dx \, (x+i\alpha) \, \ee^{-x^2/2} \ ,
\label{part-real-langevin}
\end{eqnarray}
where $x$ is a real variable.
The drift term is given by $v(x)=-\frac{1}{2x}- x$, which diverges 
at $x=0$. The probability distribution of the drift term
is only power-law suppressed at large magnitude, but the distribution
of $x$ in the thermal equilibrium approaches
$w(x)=|x|^{-1/2} \, \ee^{-x^2/2}$ as the step-size 
$\epsilon$ is reduced.

Applying the same argument to the case of the CLM, 
the FP-like equation (\ref{FP-like-eq}) 
should be obtained in the $\epsilon \rightarrow 0$ limit.
However, the $\epsilon$-expansion (\ref{OP-rewriting-P3b})
can still be subtle, and that is precisely the reason why 
the integration by parts (\ref{nec-suf-condition-old}) can 
be invalid.

\section{Demonstration of our condition}
\label{sec:examples}

In this section, we demonstrate 
our condition in section \ref{sec:criterion},
% for correct convergence
%for the validity of the $\epsilon$-expansion
which is required to justify the CLM.
For this purpose, we investigate two simple examples,
in which the CLM was thought to fail
due to the singular-drift problem and the excursion problem, respectively,
in some parameter region.
According to our new argument, however,
these failures should be attributed to the appearance of a large drift term. 
We measure the probability distribution 
%for the magnitude 
of the drift
term and show that it is only power-law suppressed at large magnitude
when the CLM fails, whereas it is exponentially suppressed when the CLM works.
Thus 
%we find that 
the failures of the CLM can be understood
in a unified manner.
Our condition is also of great practical importance since it tells us
clearly whether the obtained results are trustable or not.
%Thus our criterion can be used to tell easily and reliably
%whether the results are trustable or not.

%% justification of the CLM requires
%% that the probability for the appearance of a large drift term
%% should be suppressed exponentially.
%% %can be detected by the appearance of the large drift both 
%% %for the non-holomorphy of the drift or the large excursion.
%% For this purpose, we study two simple examples,
%% in which the CLM is known to fail in some parameter region
%% due to the singular drift problem and the excursion problem, respectively.

\subsection{A model with a singular drift}
\label{sec:model-sing}

%%%%%%%%%%%%%%%%%%%%%%%%%%%%%%%%%%%%%%%%%%%%%%%%%
\begin{figure}[t]
\centering
\includegraphics[width=7cm]{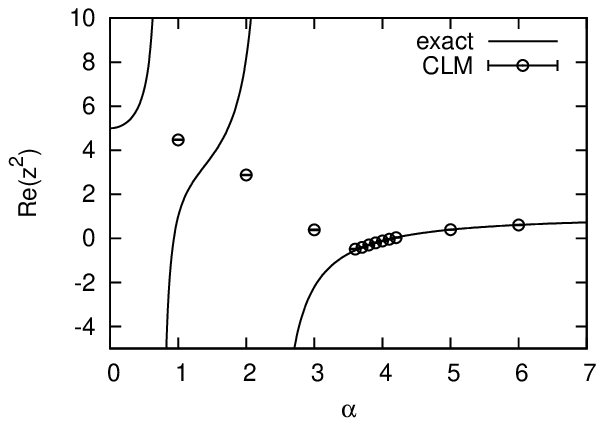}
\includegraphics[width=7cm]{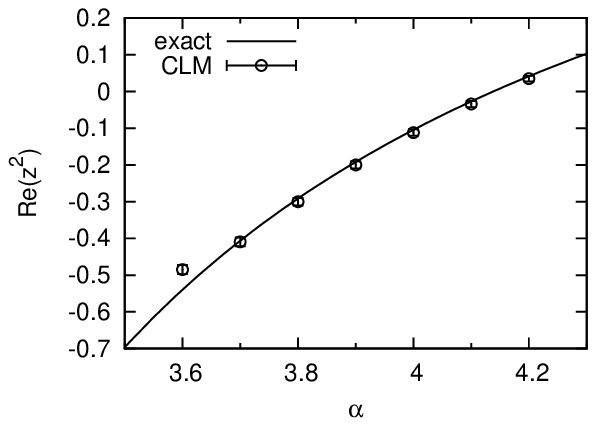}
\caption{(Left) The real part of the expectation value of ${\cal O}(z)=z^2$
obtained by the CLM for $p=4$ is plotted against $\alpha$.
The solid line represents the exact result.
(Right) Zoom-up of the same plot in the region $3.6 \le \alpha \le 4.2$.}
\label{singular_drift_result_x2}
\end{figure}
%%%%%%%%%%%%%%%%%%%%%%%%%%%%%%%%%%%%%%%%%%%%%%%%%

%%%%%%%%%%%%%%%%%%%%%%%%%%%%%%%%%%%%%%%%%%%%%%%%%%
\begin{figure}[tpb]
\centering
\includegraphics[width=7.5cm]{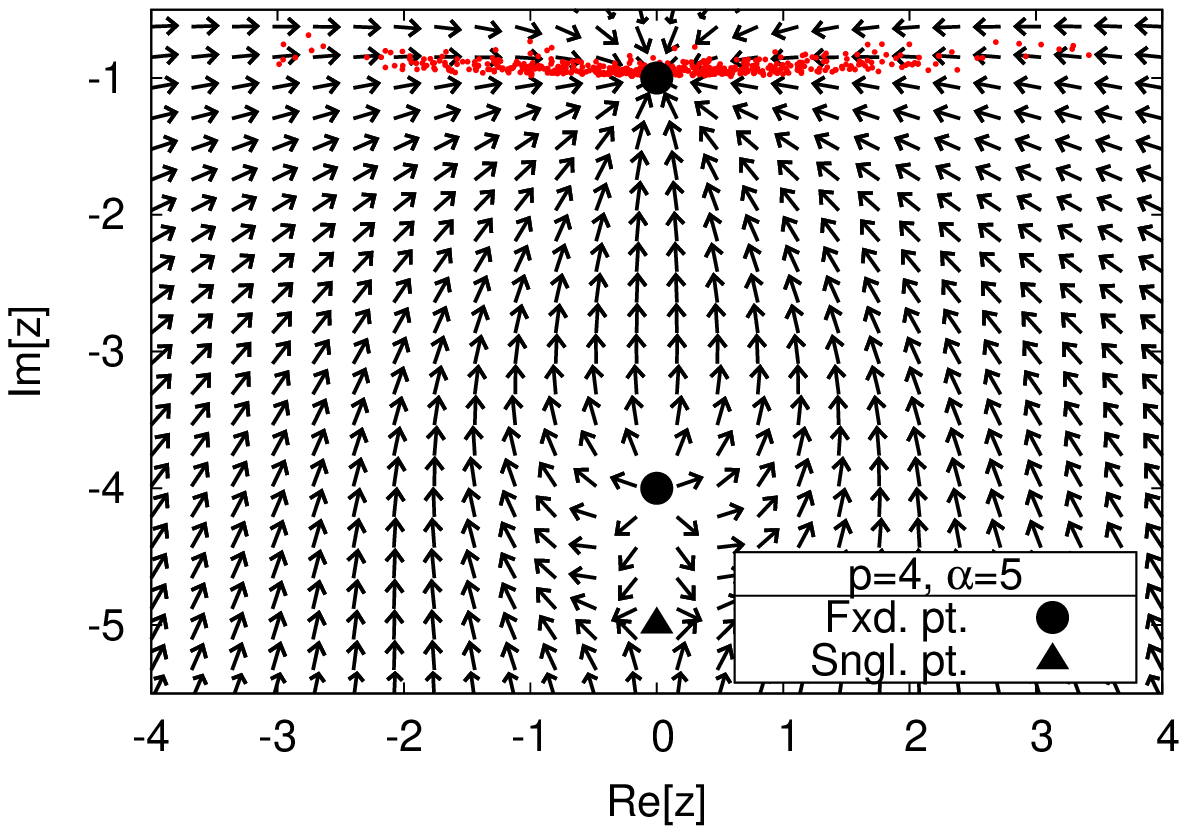}
\includegraphics[width=7.5cm]{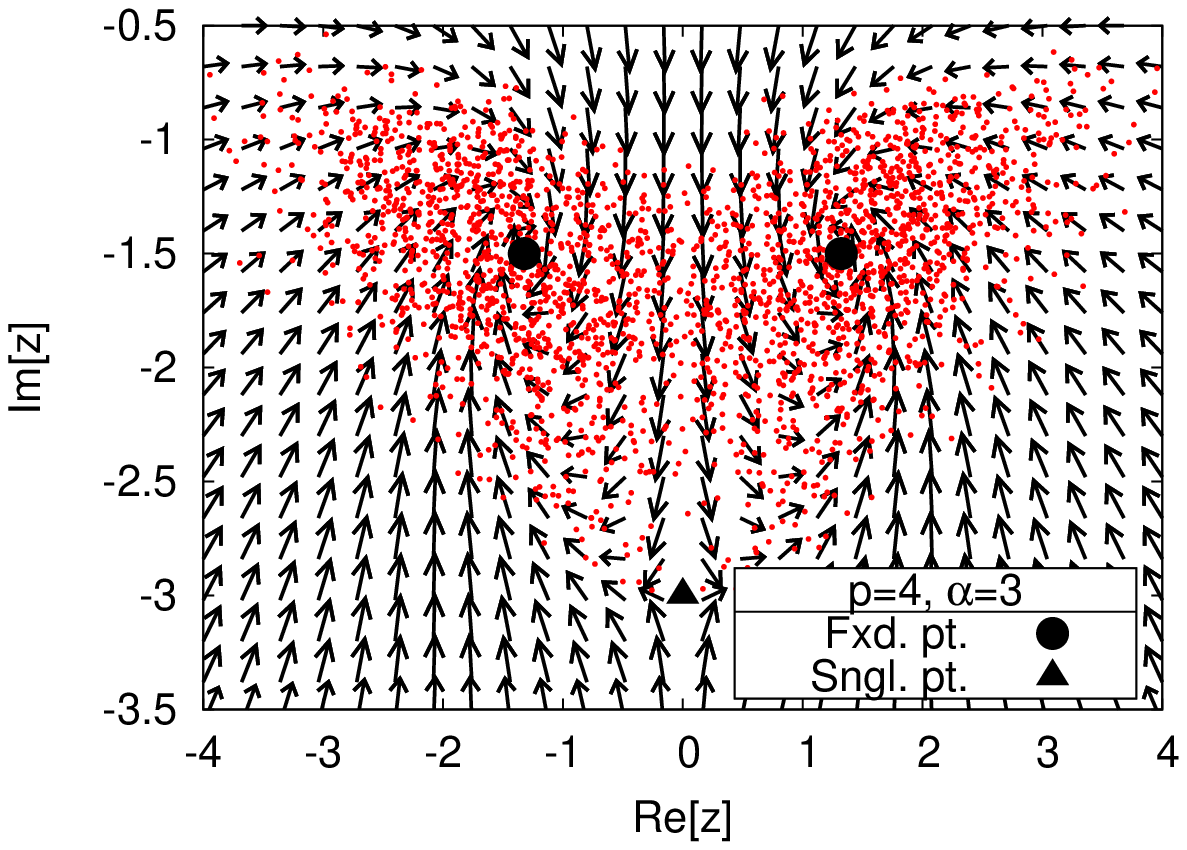}
\caption{The scatter plot of thermalized configurations (red dots)
and the flow diagram (arrows) are shown for
$\alpha =5$ (Left) and $\alpha=3$ (Right) with $p=4$. 
Filled circles represent the fixed points, and the filled triangles
represent the singular points.}
\label{classicalflow_singular_drift}
\end{figure}
%%%%%%%%%%%%%%%%%%%%%%%%%%%%%%%%%%%%%%%%%%%%%%%%%%

As a model with a singular drift, 
we consider the partition function \cite{Nishimura:2015pba}
\begin{eqnarray}
%Z = \int dx \, (x+i\alpha) \, \ee^{-(x-m)^2/2} \ ,
Z = \int dx \, w(x) \ ,\quad
w(x) = (x+i\alpha)^p \, \ee^{-x^2/2} \ ,
%Z = \int dx \, (x+i\alpha) \, \ee^{-x^2/2} \ ,
\label{part-1var}
\end{eqnarray}
where $x$ is a real variable and $\alpha$ and $p$ are real parameters.
For $\alpha \neq 0$ and $p\neq 0$, 
the weight $w(x)$ is complex, and the sign problem occurs.

We apply the CLM to \eqref{part-1var}.
Since there is no symmetry that can be used for gauge cooling,
we do not introduce the gauge cooling procedure
(\ref{eq:Langevin-discretized2-complexified-cooled0})
or the probability distribution (\ref{tilde-P}) for the transformed variables.
Otherwise, all the equations in the previous section applies
to the present case by just setting the number of variables to $N=1$.
The drift term in this model is given by
\begin{align}
v(z) =  \frac{p}{z+i\alpha} - z \ ,
\label{v-z-singular}
\end{align}
which is singular at $z=-i\alpha$.

%%%%%%%%%%%%%%%%%%%%%%%%%%%%%%%%%%%%%%%%%%%%%%%%%
\begin{figure}[tbp]
\centering
\includegraphics[width=7cm]{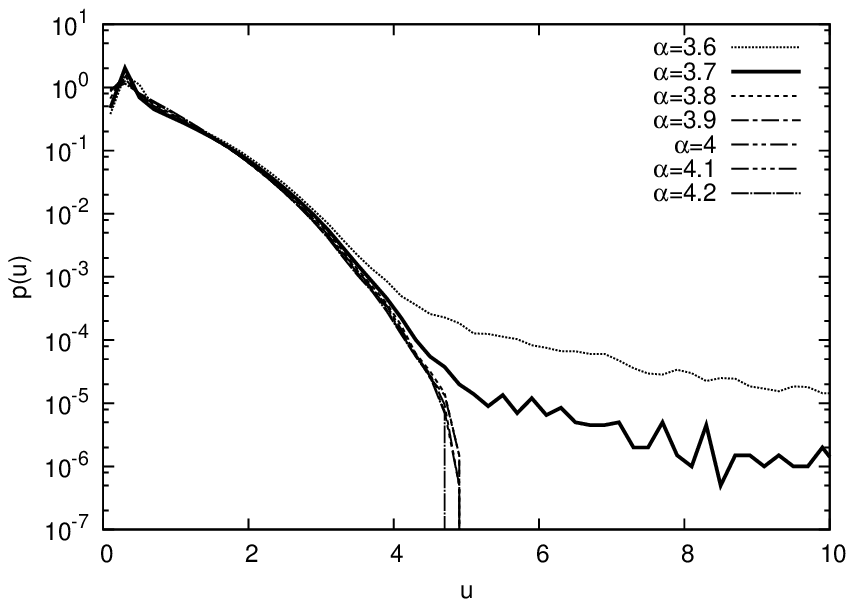}
\includegraphics[width=7cm]{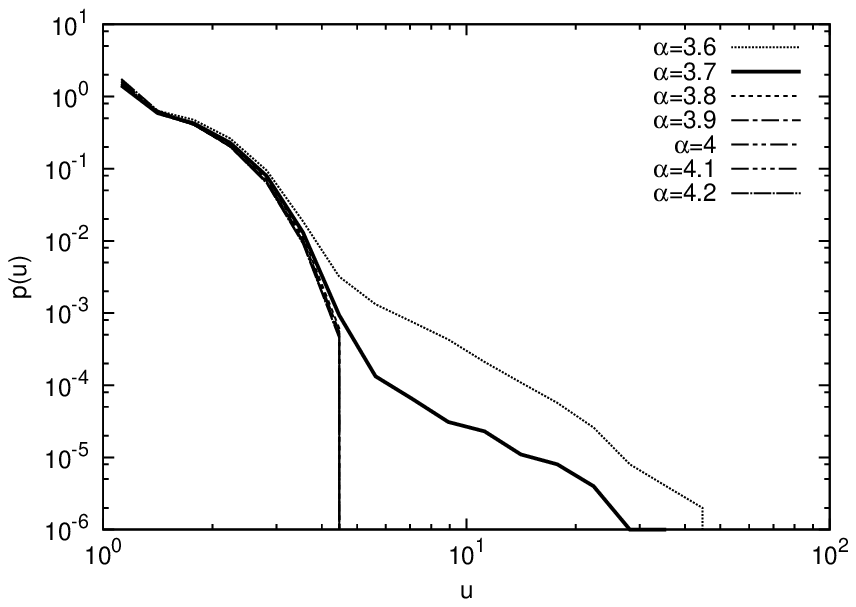}
%\includegraphics[width=7cm]{drifthist_1dv6_single_dt1e-5p4a6_dr0.5.eps}
%%\hspace{5mm}
%\includegraphics[width=7cm]{drifthist_1dv6_single_dt1e-5p4a3_dr0.5.eps}
%% \subfloat[$\alpha=6$ \label{a=6}]{\includegraphics[width=7cm]{drifthist_1dv6_single_dt1e-5p4a6_dr0.5.eps}}
%% \hspace{5mm}
%% \subfloat[$\alpha=3$ \label{a=3}]{\includegraphics[width=7cm]{drifthist_1dv6_single_dt1e-5p4a3_dr0.5.eps}}
\caption{The probability distribution $p(u)$ 
for the magnitude $u=|v|$ of the drift term 
is shown for various $\alpha$ within $3.6 \le \alpha \le 4.2$
in the semi-log (Left) and log-log (Right) plots.
%% The probability distribution $p(u)$ 
%% for the magnitude $u=|v|$ of the drift term 
%% is shown
%% for $\alpha=6$ (Left) and $\alpha=3$ (Right)
%% in a semi-log plot and a log-log plot, respectively.
%% The dashed lines are fits to
%% the straight-line behavior, which indicate 
%% exponential and power-law suppressions, respectively, at large magnitude.
%% The fitting parameters obtained by the fits are given in the Figure.
}
\label{fig_singular_hist}
\end{figure}
%%%%%%%%%%%%%%%%%%%%%%%%%%%%%%%%%%%%%%%%%%%%%%%%%

The complex Langevin simulation is performed for 
$p=4$ with various values of $\alpha$
%$\alpha=1,2,\cdots,6$
using the step-size $\epsilon=10^{-5}$.
The initial configuration is chosen to be $z=0$,
and the first $3\times 10^5$ steps are discarded for thermalization.
After that, we make $10^{10}$ steps and perform measurement every $10^3$ steps.
In Fig.~\ref{singular_drift_result_x2}
we plot the real part of the expectation value of ${\cal O}(z)=z^2$
against $\alpha$.
%as a function of $\alpha$ .
It is found that the CLM gives the correct results for $\alpha\gtrsim 3.7$.

%The reason for the change of behaviors at $\alpha = 4$ can be understood
%from the scatter plot of the configurations obtained after thermalization
%shown in Fig.~\ref{classicalflow_singular_drift}
%for $\alpha = 6$ (Left) and $\alpha=3$ (Right).

In Fig.~\ref{classicalflow_singular_drift}
we show the scatter plot of configurations obtained after thermalization
for $\alpha = 5$ (Left) and $\alpha=3$ (Right).
The data points appear near the singular point $z=-i\alpha$
for $\alpha=3$ but not for $\alpha=5$.
%This has something to do with the flow diagram
This change of behavior can be understood from the flow diagram
in the same Figure, which shows the normalized
drift term $v(z)/|v(z)|$ by an arrow at each point.
% on the complex plane for $z$.
%
%% describes the drift term (\ref{v-z-singular})
%% by an arrow representing a vector $({\rm Re}(v(z)) , {\rm Im}(v(z)))$
%% at each point on the complex plane for $z$. (The length of the
%% arrows is rescaled appropriately at each $z$, 
%% so that it is not proportional to $|v(z)|$.)
The fixed points of the flow diagram can be readily obtained by 
solving $v(z)=0$.
%For large $\alpha$ ($p < \alpha^2/4$), there are two fixed-points at 
For $\alpha >  2 \sqrt{p}$, 
there are two fixed points at 
\begin{align}
(x,y) = \left(0, - \frac{ \alpha \pm \sqrt{\alpha^2 - 4p}}{2} \right) \ , 
\end{align}
one of which ($-$) is attractive and the other ($+$) is repulsive. 
Since we adopt a real noise in the complex Langevin 
equation (\ref{eq:Langevin-discretized2-complexified}),
the thermalized configurations appear near the horizontal line
stemming from the attractive fixed point, and that is why
no configuration appears near the singular point. 

%At $p = \alpha^2/4$, the two fixed-points merge at $(0, - \alpha/2)$. 
For $\alpha =  2 \sqrt{p}$, the two fixed points merge into one at
$(0, - \alpha/2)$,
%and at small $\alpha$ ($p > \alpha^2/4$), there are two fixed-points at 
and for $\alpha <  2 \sqrt{p}$, 
there are two fixed points at 
\begin{align}
(x,y) = \left( \pm \sqrt{ p - \frac{\alpha^2}{4}}, - \frac{\alpha}{2} \right) \ ,
\end{align}
which are vortex-like.
In fact, there is a flow on the imaginary axis towards the singular point,
which makes the thermalized configurations appear near it.
%the singular point. 
Thus the property of the flow diagram changes qualitatively at
$\alpha =  2 \sqrt{p}$, which corresponds to 
$\alpha =  4$ in our case.
This is indeed close to the critical value of $\alpha$
found by comparison with the exact result 
in Fig.~\ref{singular_drift_result_x2} (Right).

%%%%%%%%%%%%%%%%%%%%%%%%%%%%%%%%%%%%%%%%%%%%%%%%%
\begin{figure}[tbp]
\centering
\includegraphics[width=7cm]{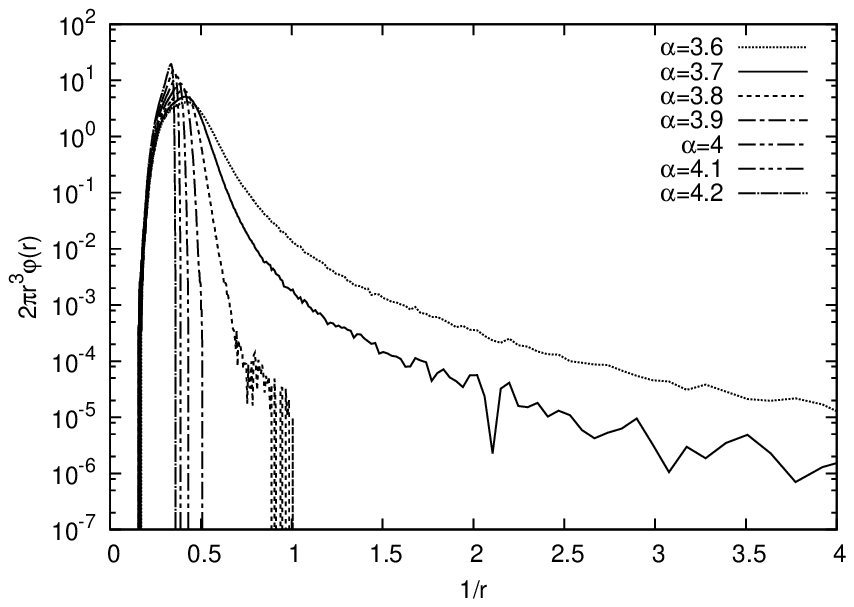}
\includegraphics[width=7cm]{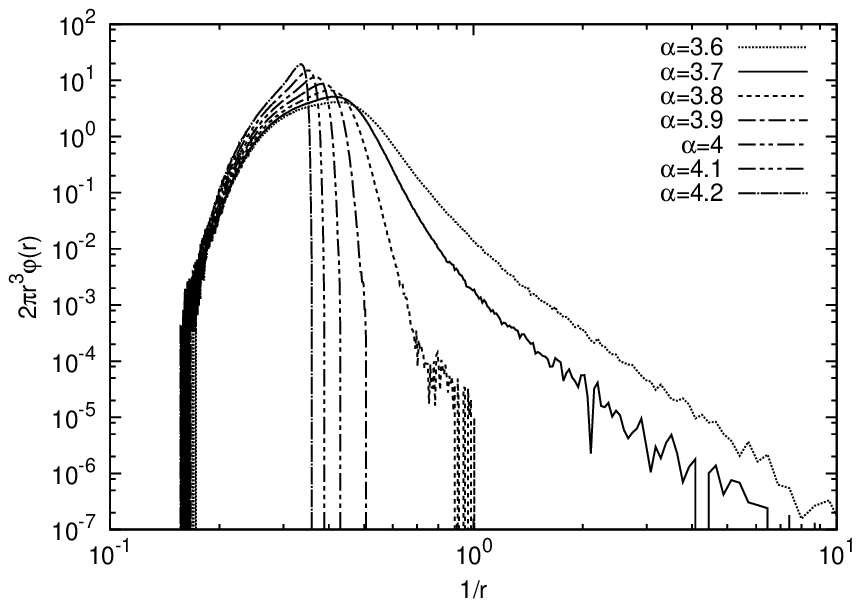}
\caption{The quantity $2 \pi r^3 \varphi(r)$,
where $\varphi(r)$ is the radial distribution defined by (\ref{def-f}),
is shown as a function of $1/r$ 
for various $\alpha$ within $3.6 \le \alpha \le 4.2$
in the semi-log (Left) and log-log (Right) plots.
}
\label{fig_singular_all}
\end{figure}
%%%%%%%%%%%%%%%%%%%%%%%%%%%%%%%%%%%%%%%%%%%%%%%%%

According to our new argument given in the previous section, 
the appearance of thermalized configurations near the singularity
of the drift term invalidates the CLM because the drift term can become
large with a probability 
%suppressed only by a power law.
that is not suppressed exponentially.
This is confirmed in Fig.~\ref{fig_singular_hist}, which
shows the probability distribution 
for the magnitude of the drift term 
for various $\alpha$ within $3.6 \le \alpha \le 4.2$
in the semi-log (Left) and log-log (Right) plots.
We find that the distribution falls off faster than exponential
for $\alpha \ge 3.8$ and that
its dependence on $\alpha$ in this region is very small.
For $\alpha \le 3.7$, the distribution follows the same behavior
as those for $\alpha \ge 3.8$ at small $u$, but it starts to
deviate from it at larger $u$. From the log-log plot,
we find that the fall-off at large $u$ is consistent with a power law.
This change of behavior occurs near the value of $\alpha$, where
the CLM starts to give wrong results as shown in 
Fig.~\ref{singular_drift_result_x2} (Right).
In fact, at $\alpha=3.7$, we cannot tell only from 
the expectation values of observables that the CLM is giving wrong results
presumably because the discrepancies are too small to be measured.
We consider this as a good feature of our condition.

In ref.~\cite{Nishimura:2015pba}, the radial distribution
\begin{eqnarray}
\varphi(r)  =  \frac{1}{2\pi r} 
\int  P(x,y,\infty)\,  \delta(\sqrt{x^2+(y+\alpha)^2}-r) \, dxdy
%\!\!  &<& \!\!\infty 
%\ ,
\label{def-f}
\end{eqnarray} 
around the singular point $(x,y)=(0,-\alpha)$ was
introduced to investigate the singular-drift problem.
Since the magnitude of the drift term is given by $u \sim 1/r$,
the probability distribution of the drift term 
is given by $p(u) \sim 2 \pi r^3 \varphi(r)$ at small $r$.
In Fig.~\ref{fig_singular_all},
we therefore show $2 \pi r^3 \varphi(r)$ as a function of 
$1/r$ in the semi-log (Left) and log-log (Right) plots.
We observe a clear power-law tail for $\alpha \le 3.7$.
Thus, the problem of the large drift term
can also be detected by the radial distribution around the singularity
if it is plotted in this way.

%% Note also that without our new argument,
%% it is not obvious how the radial distribution should fall off at small $r$
%% in order for the CLM to be justified.

\subsection{A model with a possibility of excursions}
\label{sec:model-excur}

%%%%%%%%%%%%%%%%%%%%%%%%%%%%%%%%%%%%%%%%%%%%%%%%%
\begin{figure}[t]
\centering
\includegraphics[width=7cm]{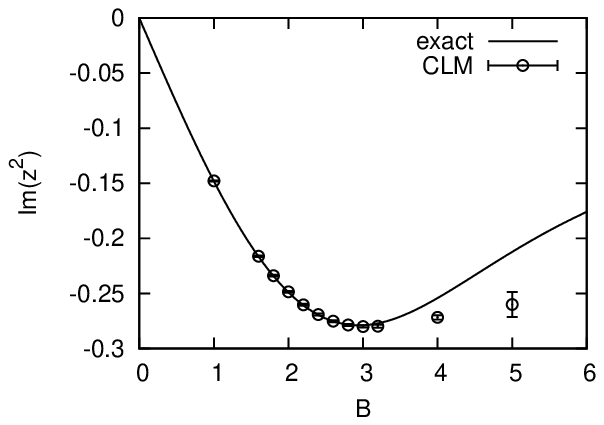}
\includegraphics[width=7cm]{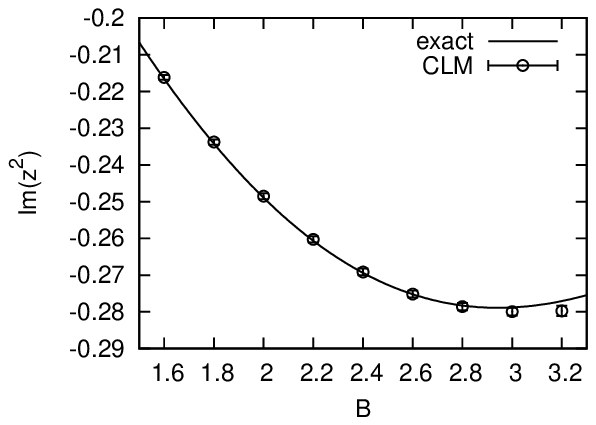}
\caption{(Left) The imaginary part of the expectation value of ${\cal O}(z)=z^2$
is plotted against $B$ for $A=1$.
The solid line represents the exact result. 
(Right) Zoom-up of the same plot in the region $1.6 \le B \le 3.2$.
}
\label{skirt_result_x2}
\end{figure}
%%%%%%%%%%%%%%%%%%%%%%%%%%%%%%%%%%%%%%%%%%%%%%%%%

%%%%%%%%%%%%%%%%%%%%%%%%%%%%%%%%%%%%%%%%%%%%%%%%%%
\begin{figure}[tp]
\centering
\includegraphics[width=7cm]{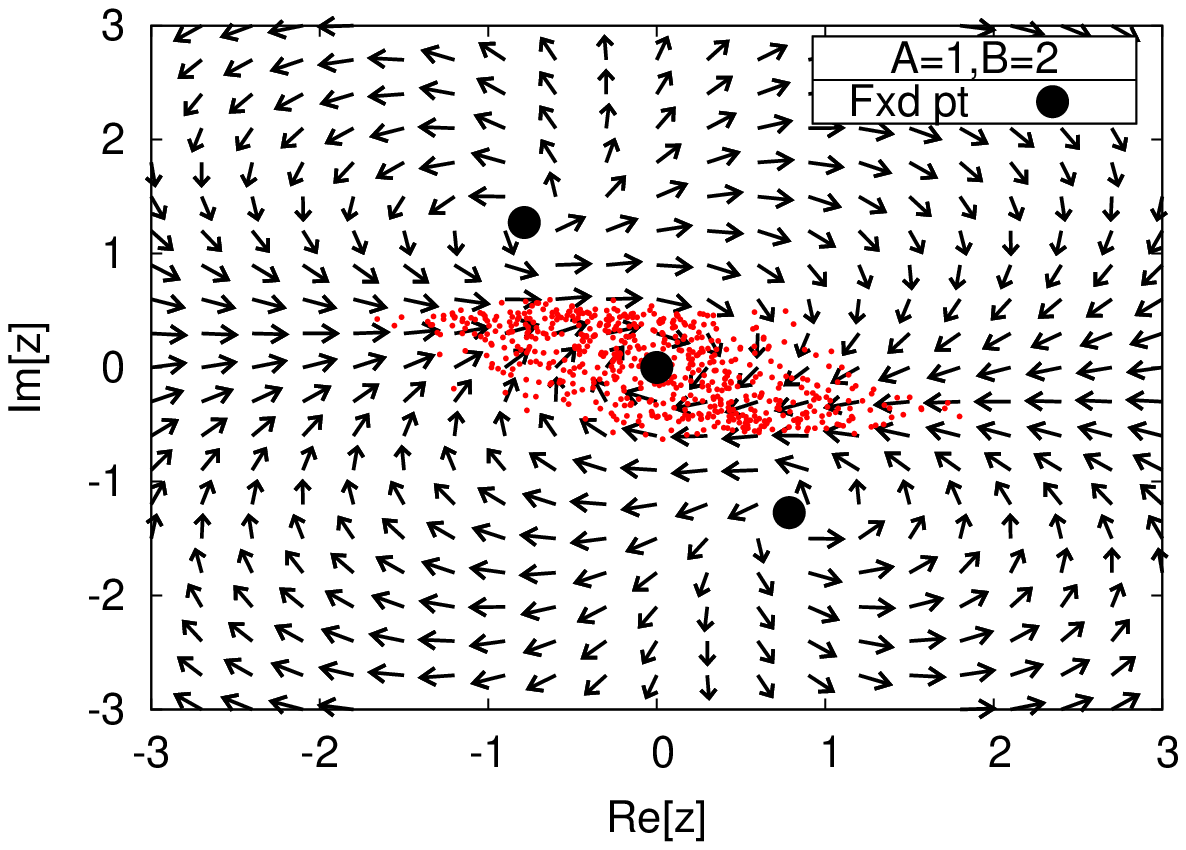}
\includegraphics[width=7cm]{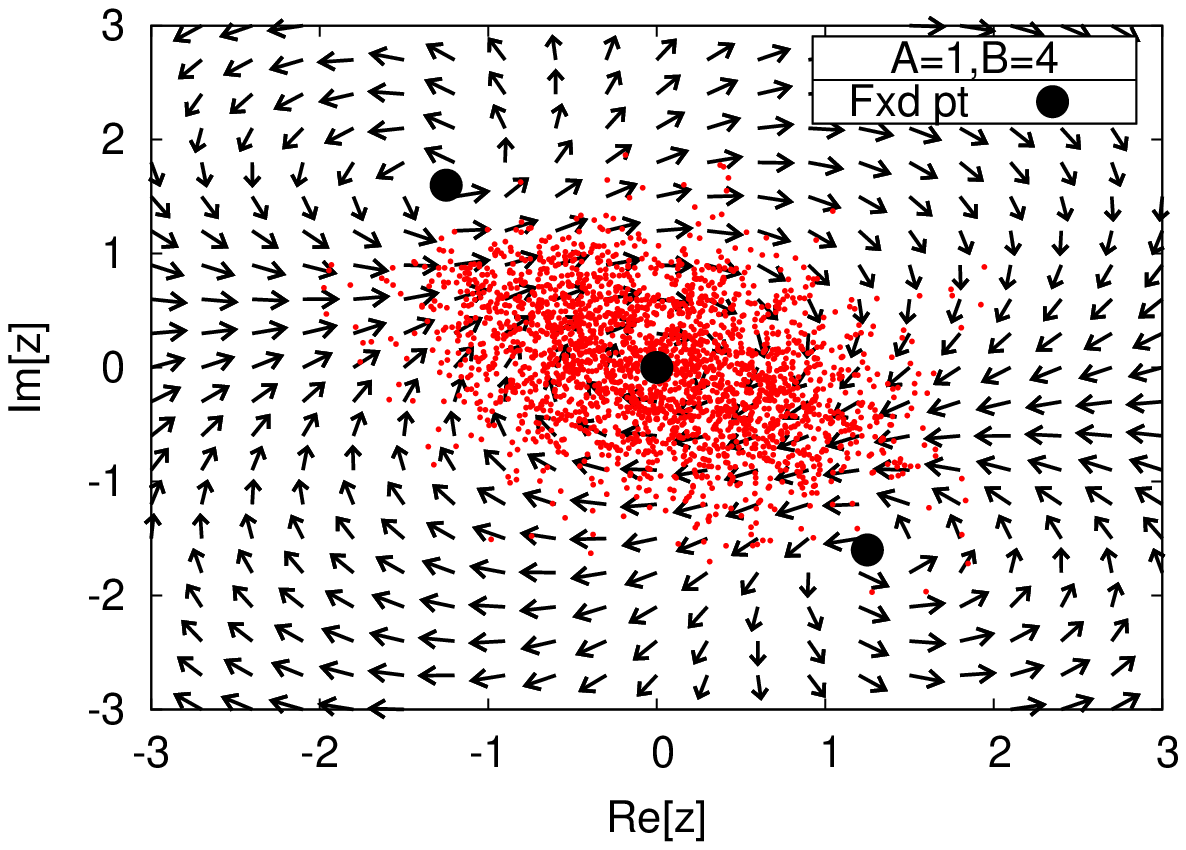}
\caption{The scatter plot of thermalized configurations (red dots)
and the flow diagram (arrows) are shown for
$B=2$ (Left) and $B=4$ (Right) with $A=1$ in both cases. 
Filled circles represent the fixed points.
There is no singular point in this model.}
\label{classicalflow_skirt}
\end{figure}
%%%%%%%%%%%%%%%%%%%%%%%%%%%%%%%%%%%%%%%%%%%%%%%%%%

As a model 
with a possibility of excursions,
we consider the partition function \cite{Aarts:2013uza}
\begin{align}
Z = \int dx \, w(x) \ ,\quad
w(x) = \ee^{-\frac{1}{2}(A+iB)x^2-\frac{1}{4}x^4} \ ,
%Z = \int dx \, (x+i\alpha) \, \ee^{-x^2/2} \ ,
\label{part-1var-2}
\end{align}
where $x$ is a real variable and $A$ and $B$ are real parameters.
For $B\neq 0$, the weight $w(x)$ is complex and the sign problem occurs.

We apply the CLM to the model \eqref{part-1var-2}.
The drift term is given by
\begin{align}
v(z) & =  - (A+iB) z - z^3 \ ,
\label{v-z-excursion}
\end{align}
which can be decomposed into the real and imaginary parts as
\begin{align}
{\rm Re} \, v(z) & =  - (Ax - By + x^3  - 3xy^2) \ ,\nonumber \\
{\rm Im} \, v(z) & =  - (Ay + Bx + 3 x^2 y - y^3 ) \ . 
\label{v-z-excursion-re-im}
\end{align}
Note that each component of the drift term can become infinitely
large with both positive and negative signs
at large $|x|$ and $|y|$,
which means that
there is a potential danger of excursions (or even runaways) in this model.
%% ${\rm Re}v(z)$ that gives a drift in the $x$-direction
%% becomes negative for sufficiently large $x$ due to the $x^3$ term,
%% and hence excursions in the real direction are suppressed.
%% On the other hand, ${\rm Im}v(z)$ that gives a drift in the $y$-direction
%% becomes positive for sufficiently large $y$ due to the $y^3$ term,
%% and there is a potential danger of excursions in the imaginary direction.

%%%%%%%%%%%%%%%%%%%%%%%%%%%%%%%%%%%%%%%%%%%%%%%%%
\begin{figure}[t]
\centering
\includegraphics[width=7cm]{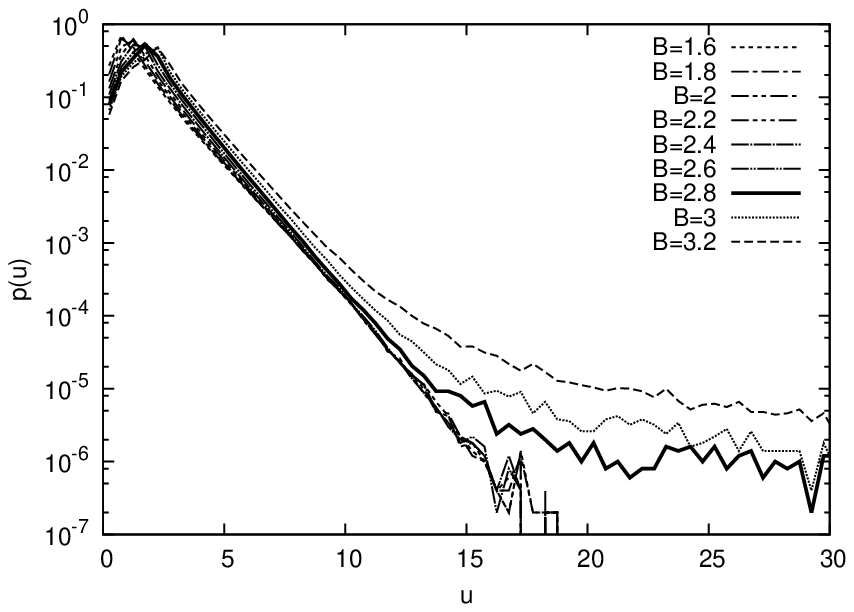}
%\includegraphics[width=7cm]{drifthist_1dv7_single_adaptive_dt1e-5A1B1_dr0.5.eps}
%\hspace{5mm}
%\includegraphics[width=7cm]{drifthist_1dv7_single_adaptive_dt1e-5A1B5_dr0.5.eps}
\includegraphics[width=7cm]{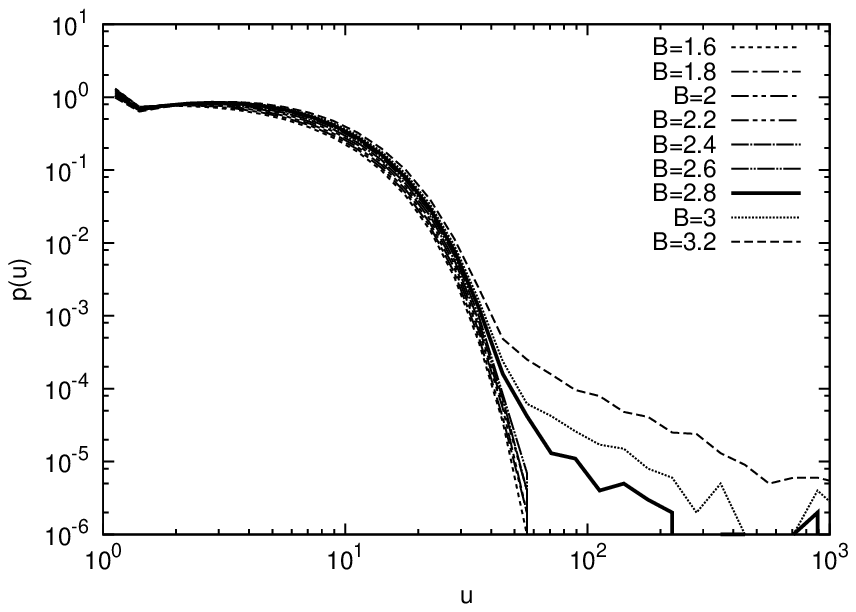}
%% \subfloat[B=1 \label{B=1}]{\includegraphics[width=7cm]{drifthist_1dv7_single_adaptive_dt1e-5A1B1_dr0.5.eps}}
%% \hspace{5mm}
%% \subfloat[B=5 \label{B=5}]{\includegraphics[width=7cm]{drifthist_1dv7_single_adaptive_dt1e-5A1B5_dr0.5.eps}}
\caption{
The probability distribution $p(u)$ 
for the magnitude $u=|v|$ of the drift term 
is shown for various $B$ within $1.6 \le B \le 3.2$
in the semi-log (Left) and log-log (Right) plots.
%% The probability distribution $p(u)$ for 
%% the magnitude $u=|v|$ of the drift term is plotted
%% for $B=1$ (Left) and $B=5$ (Right) with $A=1$ in both cases.
%% The dashed lines are fits to
%% the straight-line behavior, which indicate 
%% exponential and power-law suppressions, respectively, at large magnitude.
%% The fitting parameters obtained by the fits are given in the Figure.
}
\label{fig_skirt_hist}
\end{figure}
%%%%%%%%%%%%%%%%%%%%%%%%%%%%%%%%%%%%%%%%%%%%%%%%%

The complex Langevin simulation is performed
for $A=1$ with various values of $B$.
%$B=1,2,\cdots,5$.
The simulation parameters are the same as those in
section \ref{sec:model-sing}
%for the first example 
except that here we replace
the step-size $\epsilon=10^{-5}$ by $\epsilon=0.01/|v(z)|$
when the magnitude of the drift term $|v(z)|$ exceeds $10^3$.
The use of such an adaptive step-size \cite{Aarts:2009dg} is 
needed\footnote{The probability of
$|v(z)|$ exceeding $10^3$ is less than $10^{-4}$ even 
for the largest $B=5$ we studied.}
to avoid the runaway problem 
that occurs at $B \ge 3$.
%%
%% (See footnote \ref{foot:runaway}.)
%% \footnote{\label{foot:runaway}A runaway behavior, 
%% which was reported
%% in early works, is known to be avoidable by the use of an adaptive step-size
%% for the discretized Langevin time \cite{Aarts:2009dg}.}
%%
%The simulation results are shown 
%in Figures \ref{skirt_result_x2}, \ref{classicalflow_skirt} and \ref{fig_skirt_hist}.
In Fig.~\ref{skirt_result_x2} 
we plot the imaginary part\footnote{The real part shows similar behaviors,
but the discrepancies from the exact result at $B\gtrsim 3$ is less clear.}
of the expectation value of ${\cal O}(z)=z^2$.
We find that the CLM gives correct results for $B\lesssim 2.8$.

In Fig.~\ref{classicalflow_skirt}
we show the scatter plot of configurations obtained after thermalization
for $B=2$ (Left) and $B=4$ (Right).
The data points spread out 
in the large $|y|$ region for $B=4$
%%are actually restricted to the region 
%%$|y|\le 0.3029$ for $B=1$ \cite{Aarts:2013uza}, 
but not for $B=2$.
%This has something to do with the flow diagram
This change of behavior can be understood from the flow diagram
in the same Figure.
In fact, it was shown \cite{Aarts:2013uza} that for $B < \sqrt{3}$ ,
there is a strip-like region $|y| \le C$ in which ${\rm Im} \,v(z) \le 0 $
for $y>0$ and ${\rm Im} \, v(z) \le 0 $ for $y<0$.
%(\ref{eq:Langevin-discretized2-complexified-cooled}),
In that case, 
the thermalized configurations are strictly restricted to $|y| \le C$
as far as a real noise is used
in the complex Langevin 
equation (\ref{eq:Langevin-discretized2-complexified}).
For $B > \sqrt{3}$, this does not occur.
In fact, it was found that the distribution
in the large $|x|$ and $|y|$ region is suppressed 
only by a power law \cite{Aarts:2013uza} at sufficiently large $B$. 

According to our new argument,
this slow fall-off of the probability distribution of $x$ and $y$
invalidates the CLM 
%since the integration by parts cannot be justified.
because the drift term can become large with the probability
that is not suppressed exponentially.
This is confirmed in Fig.~\ref{fig_skirt_hist}, where
we show the probability distribution 
for the magnitude of the drift term 
for various $B$ within $1.6 \le B \le 3.2$
in the semi-log (Left) and log-log (Right) plots.
We find that the distribution falls off 
exponentially
for $B \le 2.6$ and that its 
dependence on $B$ in this region is small.
For $B \ge 2.8$, the distribution follows the same behavior
as those for $B \le 2.6$ at small $u$, but it starts to
deviate from it at larger $u$. 
From the log-log plot, we find that the
fall-off at large $u$ is consistent with a power law.
This change of behavior occurs near the value of $B$, where
the CLM starts to give wrong results as shown in 
Fig.~\ref{skirt_result_x2} (Right).
In fact, at $B=2.8$, we cannot tell only from 
the expectation values of observables that the CLM is giving wrong results
presumably because the discrepancies are too small to be measured.

\begin{figure}[tbp]
\centering
\includegraphics[width=7cm]{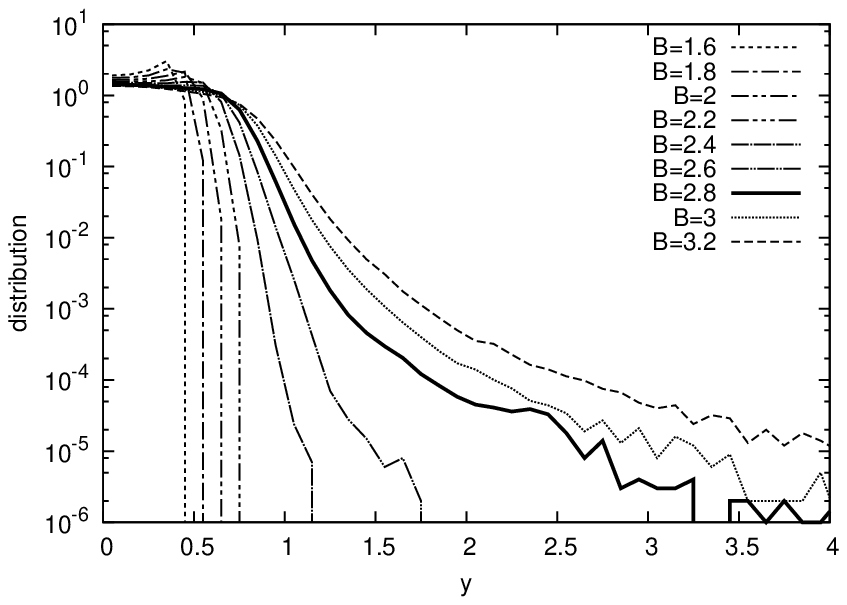}
\includegraphics[width=7cm]{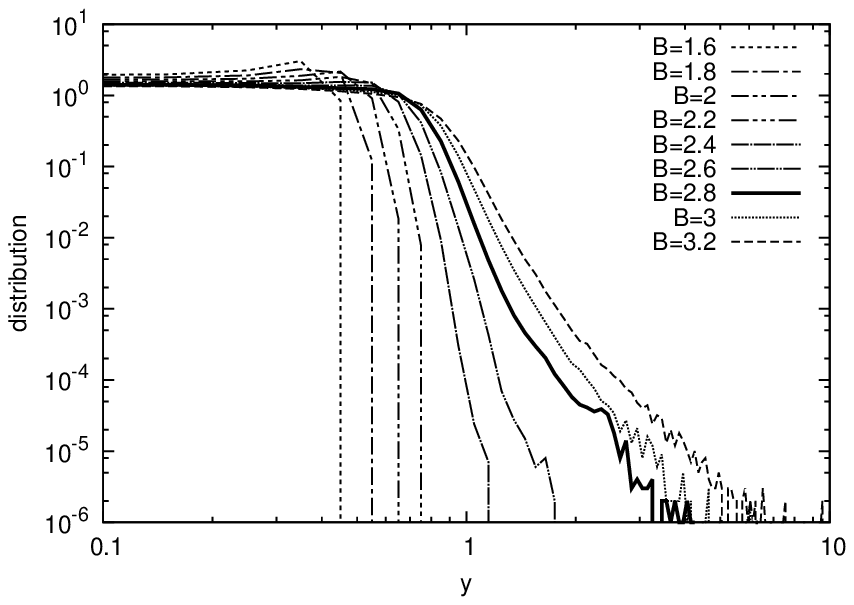}
\caption{(Left) The $y$-distribution of the thermalized 
configurations of $z=x+iy$ is shown
for various $B$ within $1.6 \le B \le 3.2$
in the semi-log (Left) and log-log (Right) plots.
}
\label{fig_excursion_all}
\end{figure}

Since the drift term is given by (\ref{v-z-excursion-re-im}) 
as a function of $x$ and $y$,
it is clear that
the large $|x|$ and large $|y|$ regions
are responsible for the slow fall-off of the
probability distribution of the drift term.
In Fig.~\ref{fig_excursion_all}, we therefore show
the $y$-distribution 
for various $B$ within $1.6 \le B \le 3.2$
in the semi-log (Left) and log-log (Right) plots.
We observe a slow fall-off consistent with a power law
%becomes prominent 
for $B \ge 2.8$.
%We confirm that the shahexponential fall-off does not occur for $B \ge 2.8$.
Thus, the problem of the large drift term
can also be detected by the $y$-distribution.
However, the change of behavior is clearer
in the probability distribution $p(u)$ for the drift term.
%albeit in a less clear way.
%% Note also that without our new argument,
%% it is not obvious how the $y$-distribution should fall off at large $y$
%% in order for the CLM to be justified.

%%%%%%%%%%%%%%%%%%%%%%%%%%%%%%%%%%%%%%%%%%%%%%%%%
%%%%%%%%%%%%%%%%%%%%%%%%%%%%%%%%%%%%%%%%%%%%%%%%%

%\section{Relaxing the assumption (\ref{lambda-def})}
%\label{sec:relaxing}

%\section*{Appendix A : Derivation of eq.~(\ref{FP-like-eq-lgt})}
%complex Langevin method 
%%%%%%%%%%%%%%%%%%%%%%%%%%%%%%%%%%%%%%%%%%%%%%%%%%%
\section{Generalization to lattice gauge theory}
\label{sec:lattice}

In this section, we discuss
the generalization of our argument in section \ref{sec:0d-model}
%complex Langevin method 
to lattice gauge theory, 
which is defined by the partition function
\begin{alignat}{3}
Z = \int dU \, w(U) = \int \prod_{n \mu} dU_{n\mu} \, w(U)  \ ,
%% Z = \int dU \, \ee^{-S(U)} = \int \prod_{n \mu} dU_{n\mu} 
%% \, \ee^{-S(U)}  \ ,
  \label{eq:part-fn-lgt}
\end{alignat}
where the weight $w(U)$ is a complex-valued function 
of the configuration $U = \{ U_{n \mu} \}$
composed of link variables $U_{n \mu} \in {\rm SU}(3)$,
and the integration measure $dU_{n\mu}$ represents the 
Haar measure for the SU(3) group.
The only complication compared with the case discussed
in section \ref{sec:0d-model}
comes from the fact that the
dynamical variables take values on a group manifold.
The Langevin equation
% for the lattice gauge theory
in such a case with a real action is discussed intensively
in refs.~\cite{Alfaro:1982ef,Drummond:1982sk,%
Guha:1982uj,Halpern:1983jt,Batrouni:1985qr}.
Using this formulation, we can easily generalize
our discussions to the case of lattice gauge theory.
In section \ref{sec:drift-term-lgt}
we discuss a new possibility
for the gauge cooling, which can reduce the magnitude of the 
drift term directly.

\subsection{The complex Langevin method}
\label{sec:CLM-lgt}

In the Langevin equation, the drift term is given by
\begin{alignat}{3}
%v_k(x) &= -  \frac{\del S(x)}{\del x_k} \ .
v_{a n\mu}(U) &= \frac{1}{w(U)} D_{a n \mu} w(U) \ ,
\label{eq:def-drift-term-lgt}
\end{alignat}
where we have defined 
the derivative operator $D_{a n \mu}$,
which acts on a function 
$f(U)$ of the unitary gauge configuration as
\begin{alignat}{3}
D_{a n \mu} f(U)
= \left. \frac{\del}{\del x} 
f(\ee^{i x t_a}  U_{n \mu}   ) \right|_{x=0}
\label{def-Dxi-lgt}
\end{alignat}
with $t_a$ being the generators of the SU(3) group
normalized by $\tr (t_a t_b)=\delta_{ab}$.
%One can generalize this formulation to 
When the weight $w(U)$ is complex, the drift term 
(\ref{eq:def-drift-term-lgt})
becomes complex, and therefore, 
the link variables evolve into 
${\rm SL}(3,\bbC)$
matrices (i.e., $3\times 3$ general complex matrices 
with the determinant one)
even if one starts from a configuration of
${\rm SU}(3)$ matrices.
Let us therefore complexify the link variables as
$U_{n \mu} \mapsto {\cal U}_{n \mu} \in  {\rm SL}(3,\bbC)$.
Then, the discretized complex Langevin equation 
is given by
\begin{alignat}{3}
{\cal U}_{n \mu}^{(\eta)} (t+\epsilon) = 
\exp \Big\{
i  \sum_a 
%\Big( - \epsilon {\cal D}_{a n \mu} S({\cal U}) 
\Big(  \epsilon \, v_{a n \mu} ({\cal U}) 
+ \sqrt{\epsilon} \, \eta_{a n \mu}(t) \Big)
\, t_a \Big\}
\,  {\cal U}_{n \mu}^{(\eta)} (t) \ ,
\label{eq:Langevin-discretized2-complexified-lgt}
\end{alignat}
where the drift term $v_{a n\mu}({\cal U})$
is obtained by analytically continuing (\ref{eq:def-drift-term-lgt}).
The probabilistic variables $\eta_{a n \mu}(t)$ are 
defined similarly to (\ref{eq:complex-noise}).
%by (\ref{def-EV-eta}).

The lattice gauge theory is invariant 
under the ${\rm SU}(3)$ gauge transformation
%% For instance, the plaquette action
%% \begin{alignat}{3}
%% S_{\rm plaquette}(U) = - \beta \sum_{n}  \sum_{\mu \neq \nu}
%% \tr (U_{n\mu} U_{n+\hat{\mu},\nu} 
%% U_{n+\hat{\nu},\mu}^{-1} U_{n\nu}^{-1} ) 
%% \label{plaquette-action}
%% \end{alignat}
%% is invariant under
\begin{alignat}{3}
U_{n \mu} ' = g_{n} \, U_{n \mu} \, g_{n+\hat{\mu}}^{-1} \ ,
\label{symmetry-lgt}
\end{alignat}
where $g_{n} \in {\rm SU}(3)$.
%% An infinitesimal transformation is denoted as
%% \begin{alignat}{3}
%% \delta U_{n \mu} = i (\lambda_{n} U_{n \mu} 
%% - U_{n \mu} \lambda_{n+\hat{\mu}} ) \ .
%% \label{symmetry-infinitesimal-lgt}
%% \end{alignat}
%% Here $\lambda_{n}$ is an element of the Lie algebra,
%% which can be expanded as
%% \begin{alignat}{3}
%% (\lambda_{n})_{jk} = \sum_a \lambda_{na} (t_a)_{jk} 
%% \label{lambda-expand-lgt}
%% \end{alignat}
%% %$\lambda = \sum_a \lambda_a t_a$
%% in terms of 
%% %Denoting 
%% the generators $t_a$ of ${\rm SU}(3)$ 
%% with real coefficients $\lambda_{na} \in \bbR$ .
%%
When one complexifies the variables 
$U_{n \mu} \mapsto {\cal U}_{n \mu} \in  {\rm SL}(3,\bbC)$,
the symmetry property of the drift term and the observables
naturally enhances to the ${\rm SL}(3,\bbC)$ gauge symmetry
that can be obtained by complexifying the original Lie group.
%% For instance, the plaquette action (\ref{plaquette-action}) becomes
%% \begin{alignat}{3}
%% S_{\rm plaquette}({\cal U}) = - \beta \sum_{n}  \sum_{\mu \neq \nu}
%% \tr ({\cal U}_{n\mu} {\cal U}_{n+ \hat{\mu},\nu} 
%% {\cal U}_{n+\hat{\nu},\mu}^{-1} {\cal U}_{n\nu}^{-1} )  \ ,
%% \label{plaquette-action-complexified}
%% \end{alignat}
%% which is invariant under
Thus, instead of (\ref{symmetry-lgt}), one obtains
\begin{alignat}{3}
{\cal U}_{n \mu} ' = g_{n} \, {\cal U}_{n \mu} \, g_{n+\hat{\mu}}^{-1}
\label{symmetry-complexified-lgt}
\end{alignat}
with $g_{n} \in {\rm SL}(3,\bbC)$.
%% An infinitesimal transformation is given by
%% \begin{alignat}{3}
%% \delta {\cal U}_{n \mu} = i (\lambda_{n} {\cal U}_{n \mu} 
%% - {\cal U}_{n \mu} \lambda_{n+\hat{\mu}} ) \ .
%% \label{symmetry-infinitesimal-complexified-lgt}
%% \end{alignat}
%% Here $\lambda$ is an element of the Lie algebra
%% for the complexified Lie group, which can be expanded
%% as (\ref{lambda-expand-lgt}) but now with complex coefficients
%% $\lambda_{na} \in \bbC$.
%% %It is important that the action is written in a holomorphic manner.

The gauge cooling \cite{Seiler:2012wz}
%(and used in all the work 
%so far \cite{Aarts:2014bwa,Sexty:2013ica,Makino:2015ooa})
modifies the complex Langevin equation 
(\ref{eq:Langevin-discretized2-complexified-lgt}) into
\begin{alignat}{3}
\widetilde{\cal U}_{n \mu}^{(\eta)} (t) & = 
g_{n} \, 
{\cal U}_{n \mu}^{(\eta)} (t) \, 
g_{n+\hat{\mu}}^{-1}   \ ,  
\label{eq:Langevin-discretized2-complexified-cooled0-lgt}
\\
{\cal U}_{n \mu}^{(\eta)} (t+\epsilon) & = 
\exp \Big\{
i \sum_a 
%\Big( - \epsilon {\cal D}_{a n \mu} S(\widetilde{\cal U}) 
\Big( \epsilon v_{a n \mu} (\widetilde{\cal U}) 
%
% {\cal D}_{a n \mu} S
%\Big|_{\widetilde{\cal U}^{(\eta)} (t)}
%
+ \sqrt{\epsilon} \eta_{a n \mu}(t) \Big)
\, t_a \Big\}
\,  \widetilde{\cal U}_{n \mu}^{(\eta)} (t)  \ ,
\label{eq:Langevin-discretized2-complexified-cooled-lgt}
\end{alignat}
where $g_{n}$ is an element of the complexified Lie group
chosen appropriately as a function of the configuration 
${\cal U}^{(\eta)}(t)$ before cooling.
We regard 
(\ref{eq:Langevin-discretized2-complexified-cooled0-lgt}) and
(\ref{eq:Langevin-discretized2-complexified-cooled-lgt})
as describing
the $t$-evolution of ${\cal U}_{n \mu}^{(\eta)}(t)$ and treat
$\widetilde{\cal U}_{n \mu}^{(\eta)} (t)$ as an intermediate object.
The basic idea is to determine $g$
in such a way that the modified Langevin process 
does not suffer from the problem of the original 
Langevin process (\ref{eq:Langevin-discretized2-complexified-lgt}).
%% The basic idea is to determine $g_{n}$
%% in such a way that the modified Langevin process 
%% (\ref{eq:Langevin-discretized2-complexified-cooled-lgt})
%% does not suffer from the problem of the original 
%% Langevin process (\ref{eq:Langevin-discretized2-complexified-lgt}).

%%%%%%%%%%%%%%%%%%%%%%%%

We consider observables ${\cal O}(U)$,
which are gauge invariant
and admit holomorphic extension to ${\cal O}({\cal U})$.
Note that the symmetry of the observables also
enhances to (\ref{symmetry-complexified-lgt}).
Its expectation value can be defined as
\begin{alignat}{3}
\Phi(t) = 
\Big\langle  {\cal O}\Big({\cal U}^{(\eta)} (t)\Big)
\Big\rangle_{\eta} 
=
\int 
d{\cal U} \, {\cal O}({\cal U}) \, P({\cal U};t) \ ,
\label{OP-rewriting-lgt}
\end{alignat}
where we have defined
the probability distribution of ${\cal U}^{(\eta)} (t)$ by
\begin{alignat}{3}
P({\cal U};t) = \Bigl\langle \prod_{n \mu}
\delta \Big({\cal U}_{n \mu} , {\cal U}_{n \mu}^{(\eta)} (t) \Big)
\Bigr \rangle_\eta \ ,
\label{def-P-xy-lgt}
\end{alignat}
using the delta function defined by
\begin{alignat}{3}
\int d  {\cal U}   \, 
f({\cal U}) \,
\delta \Big({\cal U}_{n \mu} ,  \widetilde{\cal U}_{n \mu}  \Big)
= f(\widetilde{\cal U})
\label{def-delta-lgt}
\end{alignat}
for any function $f({\cal U})$.
The integration measure $d{\cal U}$ for the complexified link variables
% that appears on the left-hand side
is given by
%represents 
the Haar measure for the ${\rm SL}(3,\bbC)$ group
normalized appropriately.
Under certain conditions, we can show that
\begin{alignat}{3}
%\lim_{\epsilon \rightarrow 0 }
\lim_{t \rightarrow \infty} 
\lim_{\epsilon \rightarrow 0 } \, 
\Phi(t)
&= 
\frac{1}{Z} \int 
d{\cal U}
%\prod_k dx_k 
\, {\cal O}({\cal U}) \, w({\cal U}) \ ,
\label{O-time-av-complex-lgt}
\end{alignat}
which implies that the CLM is justified.
%, which implies that the CLM 
%
%in the $\epsilon \rightarrow 0$ limit.
%
%Below we discuss the necessary and sufficient condition
%for this statement.
%
%Then one can show under certain conditions that

\subsection{The $t$-evolution of the expectation value}
\label{sec:t-evolution-lgt}

Let us first discuss the $t$-evolution of the expectation value $\Phi(t)$,
which is given by
%Let us therefore consider
\begin{alignat}{3}
\Phi(t + \epsilon) = 
\Big\langle  {\cal O}\Big( {\cal U}^{(\eta)} (t+\epsilon) \Big)
\Big\rangle_{\eta} 
=
\int 
d{\cal U} \, {\cal O}({\cal U}) \, P({\cal U};t+\epsilon) \ .
\label{OP-rewriting-P-lgt}
\end{alignat}
Note that the $t$-evolution of $P({\cal U};t)$ can be readily obtained from 
the complex Langevin equation 
(\ref{eq:Langevin-discretized2-complexified-cooled0-lgt}) and
(\ref{eq:Langevin-discretized2-complexified-cooled-lgt})
as\footnote{In the present case of lattice gauge theory,
we cannot perform the integration over $\eta$
explicitly as is done in the second equality of (\ref{P-evolve}).
The same comment applies also to
eqs.~(\ref{OP-rewriting-P2prev-lgt}) and (\ref{OP-rewriting-P3-lgt}).
Clearly, this is just a matter of expressions, which does not cause
any practical problems.}
\begin{alignat}{3}
P({\cal U};t+\epsilon)
=& \frac{1}{{\cal N}}\int d\eta \, 
\ee^{-\frac{1}{4}  \, 
\{ \frac{1}{N_{\rm R}} \eta_{a n \mu}(t)^{({\rm R})2}
+\frac{1}{N_{\rm I}}\eta_{a n \mu}(t)^{({\rm I})2} \} } 
\int d\widetilde{\cal U}
\nonumber \\
& \times 
\delta\Big({\cal U},
\exp \Big\{
i \sum_a 
%\Big( - \epsilon {\cal D}_{a n \mu} S(\widetilde{\cal U}) 
\Big( \epsilon v_{a n \mu} (\widetilde{\cal U}) 
+ \sqrt{\epsilon} \eta_{a n \mu}(t) \Big)
\, t_a \Big\}
\,  \widetilde{\cal U}_{n \mu}  \Big) 
\tilde{P}(\widetilde{\cal U};t)  \ ,
%% \nonumber \\
%% =& \frac{1}{\epsilon{\cal N}} \int d\tilde{x} d\tilde{y}
%% \exp \left[
%% -\left\{ 
%% \frac{\Big(x-\tilde{x}-\epsilon{\rm Re}v(\tilde{z})\Big)^2}{4 \epsilon N_{\rm R}}
%% +
%% \frac{\Big(y-\tilde{y}-\epsilon{\rm Im}v(\tilde{z})\Big)^2}{4 \epsilon N_{\rm I}}
%% \right\}
%% \right] 
%% \nonumber \\
%% & \quad \times \tilde{P}(\tilde{x},\tilde{y};t) \ ,
\label{P-evolve-lgt}
\end{alignat}
where ${\cal N}=2 \pi  \sqrt{N_{\rm R} N_{\rm I}} $
is just a normalization constant, and
we have defined the probability distribution for 
$\widetilde{\cal U}^{(\eta)}(t)$ in
(\ref{eq:Langevin-discretized2-complexified-cooled0-lgt}) as
\begin{alignat}{3}
\tilde{P}(\widetilde{\cal U};t)
&= \int d{\cal U} \,
%d\eta \, 
\delta\Big(\widetilde{\cal U},  {\cal U}^{(g)} \Big)
P({\cal U};t)  \ ,
\label{tilde-P-lgt}
\\
{\cal U}_{n \mu} ^{(g)} &= g_{n} \, {\cal U}_{n \mu} \, g_{n+\hat{\mu}}^{-1} \ .
\end{alignat}
%using we have introduced $z_k^{(g)} = g_{kl}(x,y) \, z_l$.
%%
%% We assume that this is a well-defined quantity\footnote{This assumption 
%% is violated, for instance, when $v(z)$ has a singularity at
%% $z=x_{*}+i y_{*}$,
%% and $P(x,y;t)$ is non-zero at $(x,y)=(x_{*} , y_{*})$.
%% In that case, the CLM would fail more miserably because one cannot obtain
%% a finite result. In this paper, we are concerned with a situation
%% in which one obtains a finite result, but it is wrong in the sense
%% that (\ref{O-time-av-complex}) does not hold.}.
Using (\ref{P-evolve-lgt}) in (\ref{OP-rewriting-P-lgt}),
we obtain
\begin{alignat}{3}
\Phi(t + \epsilon)  &= 
\frac{1}{{\cal N}}\int d\eta \, 
\ee^{-\frac{1}{4}  \, 
\{ \frac{1}{N_{\rm R}} \eta_{a n \mu}^{({\rm R})2}
+\frac{1}{N_{\rm I}}\eta_{a n \mu}^{({\rm I})2} \} } 
\int 
d{\cal U} \, {\cal O}({\cal U}) \, 
\int d\widetilde{\cal U}
\nonumber \\
& \times 
\delta\Big({\cal U},
\exp \Big\{
i \sum_a 
%\Big( - \epsilon {\cal D}_{a n \mu} S(\widetilde{\cal U}) 
\Big( \epsilon v_{a n \mu} (\widetilde{\cal U}) 
+ \sqrt{\epsilon} \eta_{a n \mu} \Big)
\, t_a \Big\}
\,  \widetilde{\cal U}_{n \mu} \Big) 
\tilde{P}(\widetilde{\cal U};t)  \ .
\label{OP-rewriting-P2prev-lgt}
\end{alignat}

Here we make an important assumption.
Let us note that
the convergence of the integral (\ref{OP-rewriting-lgt})
or (\ref{OP-rewriting-P2prev-lgt})
is not guaranteed because 
the observable $|{\cal O}({\cal U})|$ can become infinitely large,
and therefore it is possible that
the expectation value of ${\cal O}({\cal U})$ is ill-defined.
We restrict the observables to those
for which the integral (\ref{OP-rewriting-lgt}) converges absolutely 
at any $t\ge 0$.
%This assumption can be checked in the Langevin simulation
%by calculating the expectation value of 
%% Second, the time-evolved probability distribution
%% (\ref{P-evolve}) can be ill-defined 
%% because $|v(\tilde{z})|$ can be unbounded from above.
%% We assume that the integration of $\tilde{P}(\tilde{x},\tilde{y};t)$ 
%% over the region, in which $v(\tilde{z})$ is ill-defined,
%% is zero at any $t\ge 0$.
%% If this assumption is violated, the Langevin simulation
%% suffers from a run-away behavior, which cannot be avoided
%% by the use of an adapted step-size. (See footnote \ref{foot:runaway}.)
%%
%% In this paper, we restrict 
%% are concerned with a situation
%% in which one obtains a finite result, but it is wrong in the sense
%% that (\ref{O-time-av-complex}) does not hold.
%%
This assumption is legitimate
since we are concerned with a situation
in which one obtains a finite result, but it is wrong in the sense
that (\ref{O-time-av-complex-lgt}) does not hold.
%Therefore, it is legitimate to restrict ourselves to the
%cases in which this assumption is satisfied.
%these two assumptions are violated,
%the CLM would fail more miserably because one cannot obtain
%a finite result. 

%the CLM 
Under the above assumption, we can exchange the order of integration 
in (\ref{OP-rewriting-P2prev-lgt}) due to Fubini's theorem, and rewrite it as
\begin{alignat}{3}
\Phi(t + \epsilon)  &= 
\int 
d{\cal U} \, {\cal O}_{\epsilon}({\cal U}) \, 
\tilde{P}({\cal U};t) \ ,
\label{OP-rewriting-P2-lgt}
\end{alignat}
where we have defined
\begin{alignat}{3}
{\cal O}_{\epsilon}({\cal U}) 
&=   \frac{1}{ {\cal N}}
\int d\eta \, 
\ee^{-\frac{1}{4}  \, 
\{ \frac{1}{N_{\rm R}} \eta_k^{({\rm R})2}
+\frac{1}{N_{\rm I}}\eta_k^{({\rm I})2} \} } 
\nonumber \\
& \quad \times  {\cal O}\Big(
\exp \Big\{
i \sum_a 
%\Big( - \epsilon {\cal D}_{a n \mu} S({\cal U}) 
\Big( \epsilon v_{a n \mu} ({\cal U}) 
+ \sqrt{\epsilon} \eta_{a n \mu} \Big)
\, t_a \Big\}
\,  {\cal U}_{n \mu} 
  \Big) \ .
\label{OP-rewriting-P3-lgt}
\end{alignat}
Note that if ${\cal O}({\cal U})$ and $v_{an\mu}({\cal U})$ are holomorphic,
so is ${\cal O}_{\epsilon}({\cal U})$. When we say ``holomorphic'',
we admit the case in which the function has singular points.
%% Since these points have measure zero in the integral,
%% they do not make the integral ill-defined.
%%
%% Let us assume that we can expand the expression
%% (\ref{OP-rewriting-P2}) 
%% with respect to $\epsilon$.

In order to proceed further, we expand
(\ref{OP-rewriting-P3-lgt}) with respect to $\epsilon$ 
and perform the integration over $\eta$.
After some algebra, we get
\begin{alignat}{3}
{\cal O}_{\epsilon}({\cal U})
&= 
\mbox{\bf :} \ee^{\epsilon L} \mbox{\bf :} \, {\cal O}({\cal U})    \ ,
\label{O-t-evolve-expand-lgt}
\end{alignat}
where 
%the expression $\ee^{\epsilon L}$ is a short-hand notation for (\ref{exp-L}) and
%
%% \begin{alignat}{3}
%% \ee^{\epsilon L}
%% \equiv \sum_{n=0}^{\infty}
%%   \frac{1}{n!} \, \epsilon^n  L^n \ ,
%% \label{exp-L}
%% \end{alignat}
the operator $L$ is defined by
%The explicit form of the operator $L$ can be obtained as
\begin{alignat}{3}
L =& 
\Big(
{\rm Re} \,  v_{a n \mu} ({\cal U}) 
+ N_{\rm R} {\cal D}_{a n \mu}^{\rm (R)}
\Big)
{\cal D}_{a n \mu}^{\rm (R)}  
%\nonumber \\
%& 
+
\Big(
{\rm Im} \,  v_{a n \mu} ({\cal U}) 
+ N_{\rm I} {\cal D}_{a n \mu}^{\rm (I)}
\Big)
{\cal D}_{a n \mu}^{\rm (I)}  \ . 
\label{L-expression-lgt}
\end{alignat}
%the symbol $\top$ represents taking the dual of the operator.
In eq.~(\ref{L-expression-lgt}), we have defined
the derivative operators
\begin{alignat}{3}
{\cal D}^{\rm (R)}_{a n \mu} f({\cal U})
&= \left. \frac{\del}{\del x} 
f(\ee^{i x t_a}  {\cal U}_{n \mu}) \right|_{x=0}  \ ,
\label{def-DR-lgt} \\
{\cal D}^{\rm (I)}_{a n \mu} f({\cal U})
&= \left. \frac{\del}{\del y} 
f(\ee^{- y t_a}  {\cal U}_{n \mu}) \right|_{y=0} \ ,
\label{def-DI-lgt} 
\end{alignat}
where $f({\cal U})$ are functions 
on the complexified group manifold, which
are not necessarily holomorphic, and $x$ and $y$ 
%in eqs.~(\ref{def-DR-lgt}) and (\ref{def-DI-lgt})
are real parameters.
These derivative operators 
%(\ref{def-DR-lgt}) and (\ref{def-DI-lgt})
may be regarded as 
analogues of $\frac{\del}{\del x_k}$ and $\frac{\del}{\del y_k}$
used in section \ref{sec:0d-model}.
For later convenience, let us also define
\begin{alignat}{3}
{\cal D}_{a n \mu} &= 
\frac{1}{2}({\cal D}^{\rm (R)}_{a n \mu} - i {\cal D}^{\rm (I)}_{a n \mu})  \ ,
\label{def-D-lgt-DR-DI} \\
\bar{\cal D}_{a n \mu}   & =
\frac{1}{2}({\cal D}^{\rm (R)}_{a n \mu} + i {\cal D}^{\rm (I)}_{a n \mu}) \ ,
\label{def-Dbar-lgt}
\end{alignat}
which are analogues of 
$\frac{\del}{\del z_k}= \frac{1}{2}(\frac{\del}{\del x_k}
- i \frac{\del}{\del y_k})$ and 
$\frac{\del}{\del \bar{z}_k}= \frac{1}{2}(\frac{\del}{\del x_k}
+ i \frac{\del}{\del y_k})$, respectively.
Note that for a holomorphic function $f({\cal U})$,
we have $\bar{\cal D}_{a n \mu} f({\cal U}) = 0 $, and hence
\begin{alignat}{3}
 {\cal D}^{\rm (R)}_{a n \mu}  f({\cal U})  
 = {\cal D}_{a n \mu} f({\cal U}) \ ,
%  \nonumber \\
\quad \quad \quad 
 {\cal D}^{\rm (I)}_{a n \mu}  f({\cal U})  
 = i {\cal D}_{a n \mu} f({\cal U})  \ .
\label{holomorphic-func}
\end{alignat}
%for a holomorphic function $f({\cal U})$.
%Note that $\bar{\cal D}_{a n \mu} f({\cal U}) = 0 $
%for a holomorphic function $f({\cal U})$. 

Since ${\cal O}({\cal U})$ is a holomorphic function of ${\cal U}$, we have
\begin{alignat}{3}
%% \frac{\del f}{\del x_k} &= \frac{\del f}{\del y_k}
%% \frac{\del f}{\del y_k} &= \frac{\del f}{\del x_k}
L {\cal O} ({\cal U}) =& 
\Big(
{\rm Re} \,  v_{a n \mu} ({\cal U}) 
+ N_{\rm R} {\cal D}_{a n \mu}
\Big)
{\cal D}_{a n \mu} {\cal O}({\cal U})
 \nonumber \\
& + \Big(
{\rm Im} \, v_{a n \mu} ({\cal U})
+ i N_{\rm I} {\cal D}_{a n \mu}
\Big)
i {\cal D}_{a n \mu} {\cal O}({\cal U}) 
\nonumber \\
=& 
\Big\{  v_{a n \mu} ({\cal U})
+ (N_{\rm R}- N_{\rm I}) {\cal D}_{a n \mu}  \Big\}
{\cal D}_{a n \mu}  {\cal O}({\cal U})
\nonumber \\
=& \tilde{L} {\cal O}({\cal U}) \ ,
\label{Lf-lgt}
\end{alignat}
where we have used (\ref{NR-NI}) and defined
\begin{alignat}{3}
\tilde{L} &=
\Big( {\cal D}_{a n \mu} + v_{a n \mu} ({\cal U}) \Big)
{\cal D}_{a n \mu}  \ .
\label{L-tilde-lgt}
%\quad \quad
%\frac{\del}{\del t} {\cal O}(x;t) = L_0 {\cal O}(x;t) \ ,
\end{alignat}
Hence we can rewrite (\ref{O-t-evolve-expand-lgt}) as
\begin{alignat}{3}
{\cal O}_{\epsilon}({\cal U})
&= \mbox{\bf :} \ee^{\epsilon \tilde{L}} \mbox{\bf :} \, {\cal O}({\cal U})    \ .
\label{O-t-evolve-expand2-lgt}
\end{alignat}
%where $t = n \epsilon$.

Plugging (\ref{O-t-evolve-expand2-lgt}) in (\ref{OP-rewriting-P2-lgt}),
we formally obtain
\begin{alignat}{3}
\Phi(t + \epsilon)  &= 
\sum_{n=0}^{\infty}
  \frac{1}{n!} \, \epsilon^n  
\int 
d{\cal U} \, 
\Big( \mbox{\bf :} \tilde{L}^n  \mbox{\bf :} \,  {\cal O}({\cal U}) \Big)
\tilde{P}({\cal U};t) 
\nonumber \\
&= 
\sum_{n=0}^{\infty}
  \frac{1}{n!} \, \epsilon^n  
\int 
d{\cal U} \, 
\left.
\Big( \mbox{\bf :} \tilde{L}^n  \mbox{\bf :} \,  {\cal O}({\cal U}) \Big) 
\right|_{{\cal U}^{(g)}}
P({\cal U};t) 
\nonumber \\
&= 
\sum_{n=0}^{\infty}
  \frac{1}{n!} \, \epsilon^n  
\int 
d{\cal U} \, 
\Big( \mbox{\bf :} \tilde{L}^n  \mbox{\bf :} \,  {\cal O}({\cal U}) \Big) 
P({\cal U};t) \ .
\label{OP-rewriting-P3b-lgt}
\end{alignat}
In the third equality, we have used the fact that 
$\mbox{\bf :} \tilde{L}^n  \mbox{\bf :} \,  {\cal O}({\cal U})$ are 
invariant under the SL($3,\bbC$) transformation.
%% From (\ref{OP-rewriting-P3b})
%% it follows that $\lim_{\epsilon \rightarrow 0} \Phi(t)$ is
%% differentiable, and its derivative is given by
%% \begin{alignat}{3}
%% \frac{d}{dt} \, \lim_{\epsilon \rightarrow 0}  \Phi(t) 
%% &=  \lim_{\epsilon \rightarrow 0}  \int 
%% dx \, dy \, 
%% \Big\{ \tilde{L} \, {\cal O}(z) \Big\} \, P(x,y;t)  \ .
%% \label{OP-rewriting-P3c}
%% \end{alignat}
Thus we find \cite{Nagata:2015uga}
that the effect of the gauge cooling represented by $g$
disappears in the $t$-evolution of the SL($3,\bbC$) invariant observables,
although the $t$-evolution of the probability distribution $P({\cal U};t)$
is affected nontrivially by the gauge cooling as in (\ref{P-evolve-lgt}).

If the $\epsilon$-expansion (\ref{OP-rewriting-P3b-lgt})
is valid, we can truncate the infinite series 
for sufficiently small $\epsilon$ as
\begin{alignat}{3}
\Phi(t + \epsilon)
&= \Phi(t) + \epsilon \int 
d{\cal U} \, 
\Big\{ \tilde{L} \, {\cal O}({\cal U}) \Big\}
\, P({\cal U};t) + O(\epsilon^2) \ ,
\label{OP-rewriting-P3b-truncate-lgt}
\end{alignat}
which implies that the $\epsilon\rightarrow 0$ limit
can be taken without any problem, and we get
\begin{alignat}{3}
\frac{d}{dt} \, \Phi(t)
&=  \int d{\cal U} \, 
\Big\{ \tilde{L} \, {\cal O}({\cal U}) \Big\}
\, P({\cal U};t)  \ .
\label{OP-rewriting-P3b-cont-lim-lgt}
\end{alignat}
As we discussed in section \ref{sec:t-evolution},
eq.~(\ref{OP-rewriting-P3b-cont-lim-lgt}) 
can be violated
because of the possible breakdown of 
the expression (\ref{OP-rewriting-P3b-lgt}).
Note that the operator $\tilde{L}^n$ involves the $n$th power of
the drift term $v_{a n \mu} ({\cal U})$ in (\ref{L-tilde-lgt}),
which may become infinitely large.
Therefore, the integral that appears in (\ref{OP-rewriting-P3b-lgt})
may be divergent for large enough $n$.

\subsection{Subtlety in the use of time-evolved observables}
\label{sec:key-id-lgt}

In this section 
we assume that
the problem discussed in the previous section
does not occur and that (\ref{OP-rewriting-P3b-cont-lim-lgt}) holds.
Repeating this argument for $\tilde{L}^n \, {\cal O}({\cal U})$,
we obtain
\begin{alignat}{3}
\left( \frac{d}{dt} \right)^n \, \Phi(t)
&=  \int d{\cal U} \, 
\Big\{ \tilde{L}^n \, {\cal O}({\cal U}) \Big\}
\, P({\cal U};t)  \ .
\label{OP-rewriting-P3b-cont-lim-Ln-lgt}
\end{alignat}
Therefore, a finite time-evolution can be written
formally as
\begin{alignat}{3}
 \Phi(t+\tau)
&=  \sum_{n=0}^{\infty}
  \frac{1}{n!} \, \tau^n
\int d{\cal U} \, 
\Big\{  \tilde{L}^n \, {\cal O}({\cal U}) \Big\}
\, P({\cal U};t)  \ ,
%%  \Phi(t+\delta t) 
%% &=  \int dx \, dy \, 
%% \Big\{ \ee^{\delta t \, \tilde{L}} \, {\cal O}(z) \Big\}
%% \, P(x,y;t)  \ ,
\label{OP-rewriting-P3b-cont-lim-exp-lgt}
\end{alignat}
which is similar to (\ref{OP-rewriting-P3b-lgt}).
In order for this expression to be valid for a finite $\tau$, however,
it is not sufficient to assume that
the integral that appears in (\ref{OP-rewriting-P3b-cont-lim-exp-lgt})
is convergent for arbitrary $n$.
What matters is 
%It depends on 
the convergence radius of the 
infinite series (\ref{OP-rewriting-P3b-cont-lim-exp-lgt}).
%obtained after the integral
Below we provide
a proof of the key identity (\ref{O-time-av-complex-lgt})
assuming that
the convergence radius $\tau_{\rm conv}(t)$, which depends on $t$ in general,
is bounded from below as $\tau_{\rm conv}(t) \ge \tau_0 > 0$ 
for $0 \le t < \infty$.

In order to show (\ref{O-time-av-complex-lgt}), we first 
%consider the following lemma.
%show that 
prove the lemma
%% \footnote{In order to prove (\ref{O-time-av-complex}),
%% refs.~(\ref\cite{Aarts:2009uq,Aarts:2011ax})
%% used the time-evolved operator
%% $ {\cal O}(z;t) = \ee {t \tilde{L}} {\cal O}(z)$,
%% which has to be defined by Taylor expanding the exponential function.
%% However, it is possible that the series we get
%% $ \int dx dy \, 
%% \Big\{ \ee {t \tilde{L}}  {\cal O}(x)  \Big\} \, P(x,y;\tau)
%% = \sum_{k=0}^{\infty}
%%   \frac{1}{k!} \, t^k
%% \int dx dy \, \Big\{ (\tilde{L})^k \, {\cal O}(x+iy)  \Big\} \, P(x,y;\tau) $
%% may not converge at large $t$.
%% }
\begin{alignat}{3}
\int d{\cal U} \, \Big\{ \tilde{L}^n \, {\cal O}({\cal U}) \Big\} \, P({\cal U};t)
= \int dU \, \Big\{ (L_0)^n \, {\cal O}(U)  \Big\} \, \rho(U;t) 
\label{P-rho-rel-lgt}
\end{alignat}
for arbitrary integer $n$ and arbitrary $t \ge 0$,
where the operator $L_0$ is defined by 
\begin{alignat}{3}
%\quad \quad
%\frac{\del}{\del t} {\cal O}(x;t) = L_0 {\cal O}(x;t) \ ,
L_0 &=
%\Big(  D_{a n \mu} -D_{a n \mu} S(U) 
\Big(  D_{a n \mu}  + v_{a n \mu} (U) 
% (N_{\rm R} - N_{\rm I})
 \Big)
D_{a n \mu}  \ ,
\label{L0-expression-lgt}
\end{alignat}
and the complex valued function $\rho(U;t)$ is 
defined as the solution to 
the FP equation
\begin{alignat}{3}
\frac{\del }{\del t}\rho(U;t)
%% &= D_{a n \mu}
%% \Big\{  D_{a n \mu} S(U) + (N_{\rm R} -N_{\rm I})
%% D_{a n \mu} \Big\} \, 
%% \rho(U;t)  \nonumber \\
&= 
L_0^{\top}
\rho(U;t) 
&= 
D_{a n \mu}
\Big(  D_{a n \mu} -  v_{a n \mu} (U) \Big) \, 
\rho(U;t)  \ ,
\label{FPeq-complex-lgt}
\\
\rho(U;0)& =\rho(U) \ .
\end{alignat}
Here the symbol $L_0^{\top}$ is defined as an operator
satisfying
$\langle L_0,g \rangle=\langle f,L_0^{\top} g \rangle$,
where $\langle f, g \rangle  \equiv 
\int f(U)
g(U) dU$,
assuming that $f$ and $g$ are 
functions that allow integration by parts.
The initial condition is assumed to be
\begin{alignat}{3}
P({\cal U},;0)=\int  dU \, \rho(U;0)
\prod_{n \mu}
\delta \Big({\cal U}_{n \mu} , U_{n \mu} \Big)
\label{P-rho-initial-lgt}
\end{alignat}
%$P(x,y;0)=\rho(x,0)\ge 0$
with $\rho(U;0) \ge 0$ and $\int dU \rho(U) =1 $,
so that (\ref{P-rho-rel-lgt}) is trivially satisfied at $t=0$.

The proof of (\ref{P-rho-rel-lgt}) is then given by induction.
Let us assume that (\ref{P-rho-rel-lgt}) holds at $t=t_0$.
Then we obtain
\begin{alignat}{3}
\int d{\cal U} \, \Big\{ 
\ee^{\tau \, \tilde{L}}
 \, {\cal O}({\cal U}) \Big\} \, P({\cal U};t_0)
&= \int dU \, \Big\{ 
\ee^{\tau \, L_0 }
 \, {\cal O}(U)  \Big\} \, \rho(U;t_0) \ ,
\label{etL-lgt}
\end{alignat}
where $\tau$ should be smaller than the convergence radius of
the $\tau$-expansion (\ref{OP-rewriting-P3b-cont-lim-exp-lgt}) at $t=t_0$.
(The $\tau$-expansion on the right-hand side of (\ref{etL-lgt})
is expected to have
no problems due to the properties of
the complex weight $\rho(U;t_0)$ obtained by solving the 
FP equation (\ref{FPeq-complex-lgt}) for a well-defined system.)
%% where $\tau$ is smaller than the convergence radius, which is
%% assumed to be finite.
Since taking the derivative with respect to $\tau$ 
does not alter the convergence radius, we obtain 
\begin{alignat}{3}
\int d{\cal U} \, \Big\{ 
\ee^{\tau \, \tilde{L}}
\tilde{L}^n \, {\cal O}({\cal U}) \Big\} \, P({\cal U};t_0)
&= \int dU \,  \Big\{ 
\ee^{\tau \, L_0 }
(L_0)^n \, {\cal O}(U)  \Big\} \, \rho(U;t_0)
\label{etL-L0-lgt}
\end{alignat}
for arbitrary $n$. 
Note that
\begin{alignat}{3}
\mbox{l.h.s.\ of eq.~(\ref{etL-L0-lgt})} &=
\int d{\cal U} \, \Big\{ 
\tilde{L}^n \, {\cal O}({\cal U}) \Big\} \, P({\cal U};t_0+\tau)  \ ,
\label{P-rho-rel-2-lgt}
\end{alignat}
where we have used a relation like
(\ref{OP-rewriting-P3b-cont-lim-exp-lgt}), and
\begin{alignat}{3}
\mbox{r.h.s.\ of eq.~(\ref{etL-L0-lgt})}
&= \int dU \, \Big\{ 
(L_0)^n \, {\cal O}(U)  \Big\} \,  
\ee^{\tau \, (L_0)^\top }\rho(U;t_0) 
\nonumber \\
&= \int dU \, \Big\{ 
(L_0)^n \, {\cal O}(U)  \Big\} \,  
\rho(U;t_0+\tau)  \ ,
\label{P-rho-rel-2.5-lgt}
\end{alignat}
where we have used
integration by parts,
which is valid because the link variables $U_{n\mu}$ take values on
the compact SU(3) manifold.
In the second equality, we have used (\ref{FPeq-complex-lgt}).
Thus we find that (\ref{P-rho-rel-lgt}) holds at $t=t_0+\tau$, which
completes the proof of (\ref{P-rho-rel-lgt}) for arbitrary $t\ge 0$.

In order to show (\ref{O-time-av-complex-lgt}), we only need
to consider the $n=0$ case in (\ref{P-rho-rel-lgt}), which reads
\begin{alignat}{3}
\int d{\cal U} \,  {\cal O}({\cal U}) \, P({\cal U};t)
= \int dU \,  {\cal O}(U)   \, \rho(U;t)  \ .
\label{P-rho-rel-3-lgt}
\end{alignat}
Note that eq.~(\ref{FPeq-complex-lgt})
has a $t$-independent solution
\begin{alignat}{3}
\rho_{\rm time-indep}(U) = \frac{1}{Z} \, w(U) \ .
\label{time-indep-sol-complex-lgt}
\end{alignat}
According to the argument given 
in ref.~\cite{Nishimura:2015pba},
the solution to (\ref{FPeq-complex-lgt})
%(\ref{FPeq-complex}) 
asymptotes to
(\ref{time-indep-sol-complex-lgt}) at large $t$
if (\ref{P-rho-rel-3-lgt}) holds and $P({\cal U};t)$ converges
to a unique distribution in the $t\rightarrow \infty$ limit.
Hence, (\ref{O-time-av-complex-lgt}) follows from (\ref{P-rho-rel-3-lgt}).

\subsection{The magnitude of the drift term}
\label{sec:drift-term-lgt}

Let us consider how to define the magnitude of the drift term,
which is important in our condition for correct convergence discussed
in section \ref{sec:criterion}.
Corresponding to (\ref{def-v-magnitude}), we may define it as
\begin{alignat}{3}
%u(z) =  \max_{\vec{n}} | \vec{n} \cdot \vec{v}(z) | \ ,
u({\cal U}) =  \max_{g } \max_{a n\mu} | v_{a n\mu}({\cal U}^{(g)}) | \ ,
\label{def-v-magnitude-lgt}
\end{alignat}
where $g$ represents an ${\rm SU}(3)$ gauge transformation
(\ref{symmetry-lgt}) of the original theory.
Note that $u({\cal U})$ thus defined is invariant under (\ref{symmetry-lgt}).
This definition is not very useful, however,
because taking the maximum with respect to the gauge transformation 
is not easy to perform.
We would therefore like to propose an alternative one below,
which is similar to (\ref{def-v-magnitude-lgt}) 
but much easier to deal with.

First we note that (\ref{def-v-magnitude-lgt}) can be rewritten as
\begin{alignat}{3}
%u(z) =  \max_{\vec{n}} | \vec{n} \cdot \vec{v}(z) | \ ,
u({\cal U}) =  
\sqrt{\max_{g } \max_{ a n \mu} | v_{a n \mu}({\cal U}^{(g)}) |^2 } \ .
\label{def-v-magnitude-lgt2}
\end{alignat}
Next we replace the maximum with respect to the index $a$ by the summation
over it and define
\begin{alignat}{3}
%u(z) =  \max_{\vec{n}} | \vec{n} \cdot \vec{v}(z) | \ ,
\tilde{u}({\cal U}) &=  
%8^{-1/4}
\sqrt{\max_{g } \max_{n \mu} \sum_{a=1}^8 | v_{a n \mu}({\cal U}^{(g)}) |^2 }  
=  
%8^{-1/4}
\sqrt{ \max_{n \mu} \sum_{a=1}^8 | v_{a n \mu}({\cal U}) |^2 }  \ ,
\label{def-v-magnitude-lgt3}
\end{alignat}
where the maximum with respect to the ${\rm SU}(3)$ gauge transformation
can be omitted because the sum is gauge invariant.
%Since $8^{-1/4} u({\cal U}) \le \tilde{u}({\cal U}) \le 8^{1/4} u({\cal U})$
Since $u({\cal U}) \le \tilde{u}({\cal U}) \le 2\sqrt{2} \, u({\cal U})$
holds, $\tilde{u}({\cal U})$ may be considered a reasonable approximation to
$u({\cal U})$ for our purposes.
If the probability distribution of $\tilde{u}({\cal U})$ is suppressed
exponentially at large magnitude, so is the 
probability distribution of $u({\cal U})$, and vice versa.

The magnitude of the drift term defined by 
(\ref{def-v-magnitude-lgt}) or (\ref{def-v-magnitude-lgt3})
is not invariant under the complexified ${\rm SL}(3,\bbC)$ gauge 
transformation. Therefore, we may try to make it smaller by the gauge cooling.
In fact, the components of the drift term transform as an adjoint representation
under the gauge transformation. Namely, if we define a $3 \times 3$ matrix
$v_{n\mu}({\cal U})= \sum_{a=1}^8 v_{a n \mu}({\cal U}) \, t^a$, it transforms as
\begin{alignat}{3}
v_{n\mu}({\cal U}^{(g)})
&=   g_{n} \, v_{n\mu}({\cal U}) \, g_{n}^{-1} \ ,
\label{V-transform}
\end{alignat}
where $g_n \in {\rm SL}(3,\bbC)$.
Therefore, we can use the gauge cooling to reduce the magnitude of the drift term
associated with each site $n$ defined as
\begin{alignat}{3}
u_n ({\cal U})  &= \max _{\mu} 
\tr \Big( v_{n\mu}^\dag({\cal U}) \,  v_{n\mu}({\cal U}) \Big) \ .  
\label{drift-norm}
\end{alignat}
%using $g_n$. 
Note that this can be done site by site 
unlike the gauge cooling with the unitarity norm \cite{Seiler:2012wz},
for instance, because of the 
transformation property (\ref{V-transform}) of the drift term.

\section{Summary and discussions} 
\label{sec:conclusion}

In this paper we revisited the argument for justification of the CLM
given originally in refs.~\cite{Aarts:2009uq,Aarts:2011ax}
and extended recently to the case including 
the gauge cooling procedure in ref.~\cite{Nagata:2015uga}.
In particular, we pointed out that the use of time-evolved observables,
which are assumed to be justified for infinitely long time
in the previous argument \cite{Aarts:2009uq,Aarts:2011ax,Nagata:2015uga},
can be subtle.
In fact, we only have to use the time-evolved observables
for a finite but nonzero time
if we employ the induction with respect to the Langevin time in the argument.
This still requires that
the probability distribution
of the drift term should be suppressed, at least, exponentially 
at large magnitude.

We also clarified the condition for 
the validity of the integration by parts,
which was considered the main issue in the previous argument.
Starting with a finite step-size $\epsilon$ for the
discretized Langevin time,
we found that the integration by parts is valid
if the probability distribution
of the drift term falls off faster 
than any power law at large magnitude.
Since this is weaker than the condition obtained from 
the use of time-evolved observables for a finite time, 
we consider that the latter gives a necessary and sufficient 
condition for justifying the CLM.

%% The previous argument \cite{Aarts:2009uq,Aarts:2011ax,Nagata:2015uga}, 
%% which starts with a continuous Langevin time,
%% is clearly misleading from this point of view.
%% Note, in particular, that the 
%% integration by parts (\ref{nec-suf-condition-old})
%% is not used in our argument.

%(Of course, the observables have to be restricted to those which
%have finite expectation values in the complex Langevin process.)
%
%we have to restrict ourselves to 
%whose expectation values are finite in the CLM.)
%Thus, we arrive at the condition for correct convergence that

%the argument for justification can be completed

Our condition based on the probability distribution 
of the drift term was demonstrated
in two simple examples, in which the CLM was thought to fail
due to the singular-drift problem and the excursion problem, respectively.
We showed that 
the probability distribution is suppressed only by a power law
when the method fails, whereas it is suppressed exponentially
when the method works.
%Thus the two different problems can be understood in a unified manner 
%using our criterion.
Thus, our condition
%is useful in
provides a simple way to judge 
whether the results obtained by the method are trustable or not.
%in general applications.

The gauge cooling procedure can be included
in our argument as we did in this paper
extending our previous work \cite{Nagata:2015uga}.
Originally the gauge cooling was proposed to avoid
the excursion problem \cite{Seiler:2012wz}, and
recently it was used to solve
the singular-drift problem by adopting
different criteria for choosing the complexified gauge 
transformation \cite{Nagata:2016alq}.
Since the two problems are now understood
as the problem of a large drift term in a unified manner,
we may also choose the complexified gauge transformation
in such a way that the magnitude of the drift term is reduced.
%This idea is now being tested in the Random Matrix Theory for finite
%density QCD \cite{NNS}.
In the lattice gauge theory, such gauge cooling can be done 
site by site due to the transformation property of the drift term.
It would be interesting to see if the new type of gauge cooling,
possibly combined with the previous ones, is effective in
reducing the problem of a large drift term.

To conclude,
we consider that the present work establishes
the argument for justification of the CLM with or without gauge cooling.
The crucial point for the success of the CLM turns out to be extremely simple. 
The probability of the drift term should be suppressed exponentially
at large magnitude.
%with the probability
%suppressed only by a power law
% in order for the 
%zero step-size limit to be taken appropriately.
Now that we have such a simple understanding of the method,
we may also think of a new technique other than gauge cooling,
which enables us to enlarge the range of applicability of the CLM
further.

%%%%%%%%%%%%%%%%%%%%%%%%%%%%%%%%%%%%%%%%%%%%%%%%%%%%%%%%%%%%%%%%%%%%%%
%%%%%%%%%%%%%%%%%%%%%%%%%%%%%%%%%%%%%%%%%%%%%%%%%%%%%%%%%%%%%%%%%%%%%%
%%%%%%%%%%%%%%%%%%%%%%%%%%%%%%%%%%%%%%%%%%%%%%%%%%%%%%%%%%%%%%%%%%%%%%
\section*{Note added}
%\hspace{0.51cm}
%%%%%%%%%%%%%%%%%%%%%%%%%%%%%%%%%%%%%%%%%%%%%%%%%%%%%%%%%%%%%%%%%%%%%%

The present version of the paper has been changed significantly
from the first version put on the arXiv, where we stated
that the zero step-size limit is subtle.
Through discussions with other people,
we noticed at some point that 
this subtlety actually occurs only in the 
expression for time-evolved observables but \emph{not} in the 
Fokker-Planck-like equation. We reached this understanding after
reconsidering the case of the real Langevin method, in which 
the correct Fokker-Planck equation with a continuous Langevin time 
can be obtained even if the probability distribution 
of the drift term is suppressed only by a power law.
These points are emphasized in Sections
\ref{sec:t-evolution} and \ref{sec:real-langevin}.

%%%%%%%%%%%%%%%%%%%%%%%%%%%%%%%%%%%%%%%%%%%%%%%%%%%%%%%%%%%%%%%%%%%%%%
%%%%%%%%%%%%%%%%%%%%%%%%%%%%%%%%%%%%%%%%%%%%%%%%%%%%%%%%%%%%%%%%%%%%%% 

%%%%%%%%%%%%%%%%%%%%%%%%%%%%%%%%%%%%%%%%%%%%%%%%%%%%%%%%%%%%%%%%%%%%%%
%%%%%%%%%%%%%%%%%%%%%%%%%%%%%%%%%%%%%%%%%%%%%%%%%%%%%%%%%%%%%%%%%%%%%%
%%%%%%%%%%%%%%%%%%%%%%%%%%%%%%%%%%%%%%%%%%%%%%%%%%%%%%%%%%%%%%%%%%%%%%
\section*{Acknowledgements}
%\hspace{0.51cm}
%%%%%%%%%%%%%%%%%%%%%%%%%%%%%%%%%%%%%%%%%%%%%%%%%%%%%%%%%%%%%%%%%%%%%%
%%%%%%%%%%%%%%%%%%%%%%%%%%%%%%%%%%%%%%%%%%%%%%%%%%%%%%%%%%%%%%%%%%%%%%
%%%%%%%%%%%%%%%%%%%%%%%%%%%%%%%%%%%%%%%%%%%%%%%%%%%%%%%%%%%%%%%%%%%%%% 

The authors would like to 
%J.~N.\ 
%We 
thank J.~Bloch, K.~Fukushima, D.~Sexty and Y.~Tanizaki for valuable discussions.
K.~N.\ was supported by JSPS Grants-in-Aid for Scientific Research (Kakenhi)
Grants No.\ 26800154,  MEXT SPIRE and JICFuS.
J.~N.\ was supported in part by Grant-in-Aid 
for Scientific Research (No.\ 23244057 and 16H03988)
%and No.\ 20340048(B) for T.Y.\
from Japan Society for the Promotion of Science.
%from the Ministry of Education, 
%Science, and Culture of Japan. 
S.~S.\ was supported by the MEXT-Supported Program for the Strategic
Research Foundation at 
Private Universities ``Topological Science'' (Grant No.\ S1511006).

%\appendix

%%%%%%%%%%%%%%%%%%%%%%%%%%%%%%%%%%%%%%%%%%%%%%%%%%%%%%%%%
%%%%%%%%%%%%%%%%%%%%%%%%%%%%%%%%%%%%%%%%%%%%%%%%%%%%%%%%%


\begin{thebibliography}{3}


%\cite{Parisi:1984cs}
\bibitem{Parisi:1984cs} 
  G.~Parisi,
\emph{On complex probabilities},
  Phys.\ Lett.\ B {\bf 131} (1983) 393.
  %%CITATION = PHLTA,B131,393;%%
  %134 citations counted in INSPIRE as of 24 Mar 2015
%\cite{Klauder:1983sp}
\bibitem{Klauder:1983sp}
  J.~R.~Klauder,
\emph{Coherent state Langevin equations for canonical quantum systems 
with applications to the quantized Hall effect},
  Phys.\ Rev.\ A {\bf 29} (1984) 2036.
  %%CITATION = PHRVA,A29,2036;%%
  %64 citations counted in INSPIRE as of 24 mar 2015


%\cite{Aarts:2014bwa}
\bibitem{Aarts:2014bwa} 
  G.~Aarts, E.~Seiler, D.~Sexty and I.~O.~Stamatescu,
\emph{Simulating QCD at nonzero baryon density to all orders 
in the hopping parameter expansion},
  Phys.\ Rev.\ D {\bf 90} (2014) no. 11, 114505 
  [{\tt arXiv:1408.3770 [hep-lat]}].
  %%CITATION = ARXIV:1408.3770;%%
  %12 citations counted in INSPIRE as of 24 mar 2015


%\cite{Sexty:2013ica}
\bibitem{Sexty:2013ica} 
  D.~Sexty,
\emph{Simulating full QCD at nonzero density using the complex Langevin equation},
  Phys.\ Lett.\ B {\bf 729} (2014) 108 
  [{\tt arXiv:1307.7748 [hep-lat]}].
  %%CITATION = ARXIV:1307.7748;%%
  %40 citations counted in INSPIRE as of 24 mar 2015

%\cite{Seiler:2012wz}
\bibitem{Seiler:2012wz} 
  E.~Seiler, D.~Sexty and I.~O.~Stamatescu,
\emph{Gauge cooling in complex Langevin for QCD with heavy quarks},
  Phys.\ Lett.\ B {\bf 723} (2013) 213 
  [{\tt arXiv:1211.3709 [hep-lat]}].
  %%CITATION = ARXIV:1211.3709;%%
  %33 citations counted in INSPIRE as of 24 Mar 2015


%\cite{Mollgaard:2013qra}
\bibitem{Mollgaard:2013qra} 
  A.~Mollgaard and K.~Splittorff,
\emph{Complex Langevin dynamics for chiral Random Matrix Theory},
  Phys.\ Rev.\ D {\bf 88} (2013) no. 11, 116007 
  [{\tt arXiv:1309.4335 [hep-lat]}].
  %%CITATION = ARXIV:1309.4335;%%
  %19 citations 

%\cite{Mollgaard:2014mga}
\bibitem{Mollgaard:2014mga} 
  A.~Mollgaard and K.~Splittorff,
\emph{Full simulation of chiral random matrix theory 
at nonzero chemical potential by complex Langevin},
  Phys.\ Rev.\ D {\bf 91} (2015) no. 3, 036007 
  [{\tt arXiv:1412.2729 [hep-lat]}].
  %%CITATION = ARXIV:1412.2729;%%
  %3 citations counted in INSPIRE as of 24 Apr 2015


%\cite{Fodor:2015doa}
\bibitem{Fodor:2015doa}
  Z.~Fodor, S.~D.~Katz, D.~Sexty and C.~T\"or\"ok,
\emph{Complex Langevin dynamics for dynamical QCD 
at nonzero chemical potential: a comparison with multiparameter reweighting},
  Phys.\ Rev.\ D {\bf 92} (2015) no.9, 094516
%  doi:10.1103/PhysRevD.92.094516
  [{\tt arXiv:1508.05260 [hep-lat]}].
  %%CITATION = doi:10.1103/PhysRevD.92.094516;%%
  %11 citations counted in INSPIRE as of 21 Jun 2016


%\cite{Sinclair:2015kva}
\bibitem{Sinclair:2015kva} 
  D.~K.~Sinclair and J.~B.~Kogut,
\emph{Exploring complex-Langevin methods for finite-density QCD},
  arXiv:1510.06367 [hep-lat].
  %%CITATION = ARXIV:1510.06367;%%




%\cite{Nagata:2016alq}
\bibitem{Nagata:2016alq}
  K.~Nagata, J.~Nishimura and S.~Shimasaki,
\emph{Gauge cooling for the singular-drift problem 
in the complex Langevin method: 
a test in Random Matrix Theory for finite density QCD},
  arXiv:1604.07717 [hep-lat].
  %%CITATION = ARXIV:1604.07717;%%


%\cite{Parisi:1980ys}
\bibitem{Parisi:1980ys} 
  G.~Parisi and Y.~s.~Wu,
\emph{Perturbation theory without gauge fixing},
  Sci.\ Sin.\  {\bf 24} (1981) 483.
  %%CITATION = SSINA,24,483;%%
  %574 citations counted in INSPIRE as of 24 Mar 2015

%\cite{Damgaard:1987rr}
\bibitem{Damgaard:1987rr} 
  P.~H.~Damgaard and H.~Huffel,
\emph{Stochastic Quantization},
  Phys.\ Rept.\ {\bf 152} (1987) 227.
  %%CITATION = PRPLC,152,227;%%
  %203 citations counted in INSPIRE as of 24 Mar 2015



%% \bibitem{Aarts:2009dg}
%% G.~Aarts, F.~A. James, E.~Seiler, and I.-O. Stamatescu, 
%% \emph{Adaptive stepsize and instabilities in complex Langevin dynamics},
%% Phys.\ Lett.\ {\bf B687} (2010) 154 [{\tt arXiv:0912.0617}].


%\cite{Aarts:2009uq}
\bibitem{Aarts:2009uq} 
  G.~Aarts, E.~Seiler and I.~O.~Stamatescu,
\emph{The complex Langevin method: when can it be trusted?},
  Phys.\ Rev.\ D {\bf 81} (2010) 054508 
  [{\tt arXiv:0912.3360 [hep-lat]}].
  %%CITATION = ARXIV:0912.3360;%%
  %65 citations counted in INSPIRE as of 10 Aug 2015


%\cite{Aarts:2011ax}
\bibitem{Aarts:2011ax} 
  G.~Aarts, F.~A.~James, E.~Seiler and I.~O.~Stamatescu,
\emph{Complex Langevin: etiology and diagnostics of its main problem},
  Eur.\ Phys.\ J.\ C {\bf 71} (2011) 1756 
  [{\tt arXiv:1101.3270 [hep-lat]}].
  %%CITATION = ARXIV:1101.3270;%%
  %43 citations counted in INSPIRE as of 24 Mar 2015



%\cite{Nishimura:2015pba}
\bibitem{Nishimura:2015pba}
  J.~Nishimura and S.~Shimasaki,
\emph{New insights into the problem with a singular drift term 
in the complex Langevin method},
  Phys.\ Rev.\ D {\bf 92} (2015) 1, 011501
  [{\tt arXiv:1504.08359 [hep-lat]}].
  %%CITATION = ARXIV:1504.08359;%%
  %2 citations counted in INSPIRE as of 29 Jul 2015


%\cite{Aarts:2013uxa}
\bibitem{Aarts:2013uxa} 
  G.~Aarts, L.~Bongiovanni, E.~Seiler, D.~Sexty and I.~O.~Stamatescu,
\emph{Controlling complex Langevin dynamics at finite density},
  Eur.\ Phys.\ J.\ A {\bf 49} (2013) 89 
  [{\tt arXiv:1303.6425 [hep-lat]}].
  %%CITATION = ARXIV:1303.6425;%%
  %41 citations counted in INSPIRE as of 13 May 2015


%\cite{Aarts:2016qrv}
\bibitem{Aarts:2016qrv}
  G.~Aarts, F.~Attanasio, B.~J\"ager a D.~Sexty,
\emph{The QCD phase diagram in the limit of heavy quarks 
using complex Langevin dynamics},
arXiv:1606.05561 [hep-lat].
  %%CITATION = ARXIV:1606.05561;%%


%\cite{Nagata:2015uga}
\bibitem{Nagata:2015uga} 
  K.~Nagata, J.~Nishimura and S.~Shimasaki,
\emph{Justification of the complex Langevin method 
with the gauge cooling procedure},
PTEP {\bf 2016} (2016) no.1, 013B01
[{\tt arXiv:1508.02377 [hep-lat]}].
  %%CITATION = ARXIV:1508.02377;%%
  %5 citations counted in INSPIRE as of 24 Oct 2015

%\cite{Aarts:2013uza}
\bibitem{Aarts:2013uza} 
  G.~Aarts, P.~Giudice and E.~Seiler,
\emph{Localised distributions and criteria for correctness 
in complex Langevin dynamics},
  Annals Phys.\  {\bf 337} (2013) 238
%  doi:10.1016/j.aop.2013.06.019
 [{\tt arXiv:1306.3075 [hep-lat]}].
  %%CITATION = doi:10.1016/j.aop.2013.06.019;%%
  %22 citations counted in INSPIRE as of 14 Apr 2016


%% %\cite{Makino:2015ooa}
%% \bibitem{Makino:2015ooa}
%%   H.~Makino, H.~Suzuki and D.~Takeda,
%%   %``Complex Langevin method applied to the 2D $SU(2)$ Yang--Mills theory,''
%%   arXiv:1503.00417 [hep-lat].
%%   %%CITATION = ARXIV:1503.00417;%%


%\cite{Greensite:2014cxa}
\bibitem{Greensite:2014cxa} 
  J.~Greensite,
\emph{Comparison of complex Langevin and mean field methods 
applied to effective Polyakov line models},
  Phys.\ Rev.\ D {\bf 90} (2014) no. 11, 114507 
  [{\tt arXiv:1406.4558 [hep-lat]}].
  %%CITATION = ARXIV:1406.4558;%%
  %8 citations counted in INSPIRE as of 24 Mar 2015


%% \bibitem{Seiler}
%% E.~Seiler,
%% \emph{Langevin with meromorphic drift: problems and partial solutions},
%% lecture at EMMI Workshop: SIGN 2014, 18-21 February 2014.


%% %\cite{Aarts:2014nxa}
%% \bibitem{Aarts:2014nxa} 
%%   G.~Aarts, L.~Bongiovanni, E.~Seiler and D.~Sexty,
%%   %``Some remarks on Lefschetz thimbles and complex Langevin dynamics,''
%%   JHEP {\bf 1410}, 159 (2014)
%%   [arXiv:1407.2090 [hep-lat]].
%%   %%CITATION = ARXIV:1407.2090;%%
%%   %11 citations counted in INSPIRE as of 24 mar 2015





%-------------------------------


%\cite{Witten:2010cx}
\bibitem{Witten:2010cx}
  E.~Witten,
\emph{Analytic continuation of Chern-Simons theory},
  arXiv:1001.2933 [hep-th].
  %%CITATION = ARXIV:1001.2933;%%
  %120 citations counted in INSPIRE as of 03 août 2015


%\cite{Cristoforetti:2012su}
\bibitem{Cristoforetti:2012su}
%  M.~Cristoforetti {\it et al.} [AuroraScience Collaboration],
M.~Cristoforetti, F.~Di Renzo and L.~Scorzato,
\emph{New approach to the sign problem in quantum field theories: 
high density QCD on a Lefschetz thimble},
  Phys.\ Rev.\ D {\bf 86} (2012) 074506 
  [{\tt arXiv:1205.3996 [hep-lat]}].
  %%CITATION = ARXIV:1205.3996;%%
  %58 citations counted in INSPIRE as of 03 août 2015

%\cite{Cristoforetti:2013wha}
\bibitem{Cristoforetti:2013wha} 
  M.~Cristoforetti, F.~Di Renzo, A.~Mukherjee and L.~Scorzato,
\emph{Monte Carlo simulations on the Lefschetz thimble: taming the sign problem},
  Phys.\ Rev.\ D {\bf 88} (2013) no. 5, 051501 
  [{\tt arXiv:1303.7204 [hep-lat]}].
  %%CITATION = ARXIV:1303.7204;%%
  %24 citations counted in INSPIRE as of 10 Aug 2015


%\cite{Fujii:2013sra}
\bibitem{Fujii:2013sra}
  H.~Fujii, D.~Honda, M.~Kato, Y.~Kikukawa, S.~Komatsu and T.~Sano,
\emph{Hybrid Monte Carlo on Lefschetz thimbles: a 
study of the residual sign problem},
  JHEP {\bf 1310} (2013) 147 
  [{\tt arXiv:1309.4371 [hep-lat]}].
  %%CITATION = ARXIV:1309.4371;%%
  %30 citations counted in INSPIRE as of 03 Aug 2015

%\cite{Mukherjee:2014hsa}
\bibitem{Mukherjee:2014hsa} 
  A.~Mukherjee and M.~Cristoforetti,
\emph{Lefschetz thimble Monte Carlo for many-body theories: a Hubbard model study},
  Phys.\ Rev.\ B {\bf 90} (2014) no. 3, 035134 
  [{\tt arXiv:1403.5680 [cond-mat.str-el]}].
  %%CITATION = ARXIV:1403.5680;%%
  %7 citations counted in INSPIRE as of 10 Aug 2015



%\cite{DiRenzo:2015foa}
\bibitem{DiRenzo:2015foa}
  F.~Di Renzo and G.~Eruzzi,
\emph{Thimble regularization at work: 
from toy models to chiral random matrix theories},
  Phys.\ Rev.\ D {\bf 92} (2015) no.8, 085030
%  doi:10.1103/PhysRevD.92.085030
  [{\tt arXiv:1507.03858 [hep-lat]}].
  %%CITATION = doi:10.1103/PhysRevD.92.085030;%%
  %11 citations counted in INSPIRE as of 21 Jun 2016


%\cite{Fukushima:2015qza}
\bibitem{Fukushima:2015qza}
  K.~Fukushima and Y.~Tanizaki,
\emph{Hamilton dynamics for the Lefschetz thimble integration 
akin to the complex Langevin method},
  %``Hamilton dynamics for Lefschetz-thimble integration 
%akin to the complex Langevin method,''
  PTEP {\bf 2015} (2015) no.11,  111A01
%  doi:10.1093/ptep/ptv152
  [{\tt arXiv:1507.07351 [hep-th]}].
  %%CITATION = doi:10.1093/ptep/ptv152;%%
  %10 citations counted in INSPIRE as of 21 Jun 2016


%\cite{Alexandru:2015sua}
\bibitem{Alexandru:2015sua}
  A.~Alexandru, G.~Basar, P.~F.~Bedaque, G.~W.~Ridgway and N.~C.~Warrington,
\emph{Sign problem and Monte Carlo calculations beyond Lefschetz thimbles},
  JHEP {\bf 1605} (2016) 053
%  doi:10.1007/JHEP05(2016)053
  [{\tt arXiv:1512.08764 [hep-lat]}].
  %%CITATION = doi:10.1007/JHEP05(2016)053;%%
  %3 citations counted in INSPIRE as of 21 Jun 2016

%-------------------------------

%\bibitem{NNS}
% K.~Nagata, J.~Nishimura and S.~Shimasaki, work in progress.

\bibitem{Aarts:2009dg}
G.~Aarts, F.~A. James, E.~Seiler, and I.-O. Stamatescu, 
\emph{Adaptive stepsize and instabilities in complex Langevin dynamics},
Phys.\ Lett.\ {\bf B687} (2010) 154 
[{\tt arXiv:0912.0617}].

%\cite{Duncan:2012tc}
\bibitem{Duncan:2012tc}
  A.~Duncan and M.~Niedermaier,
\emph{Temporal breakdown and Borel resummation in the complex Langevin method},
Annals Phys.\  {\bf 329} (2013) 93
%  doi:10.1016/j.aop.2012.09.011
[{\tt arXiv:1205.0307 [quant-ph]}].
  %%CITATION = doi:10.1016/j.aop.2012.09.011;%%
  %10 citations counted in INSPIRE as of 15 Jul 2016



%% %\cite{Ambjorn:1985iw}
%% \bibitem{Ambjorn:1985iw}
%%   J.~Ambjorn and S.~K.~Yang,
%%   %``Numerical Problems in Applying the Langevin Equation to Complex Effective Actions,''
%%   Phys.\ Lett.\ B {\bf 165}, 140 (1985).
%%   %%CITATION = PHLTA,B165,140;%%
%%   %62 citations counted in INSPIRE as of 04 août 2015

%\cite{Alfaro:1982ef}
\bibitem{Alfaro:1982ef} 
  J.~Alfaro and B.~Sakita,
\emph{Stochastic quantization and large N limit of U(N) gauge theory},
  CCNY-HEP-82-16, C82-09-09.1.
  %%CITATION = CCNY-HEP-82-16, C82-09-09.1;%%

%\cite{Drummond:1982sk}
\bibitem{Drummond:1982sk}
  I.~T.~Drummond, S.~Duane and R.~R.~Horgan,
\emph{The stochastic method for numerical simulations: higher order corrections},
  Nucl.\ Phys.\ B {\bf 220} (1983) 119.
  %%CITATION = NUPHA,B220,119;%%
  %85 citations counted in INSPIRE as of 20 May 2015


%\cite{Guha:1982uj}
\bibitem{Guha:1982uj} 
  A.~Guha and S.~C.~Lee,
\emph{Stochastic quantization of matrix and lattice gauge models},
  Phys.\ Rev.\ D {\bf 27} (1983) 2412.
  %%CITATION = PHRVA,D27,2412;%%
  %52 citations counted in INSPIRE as of 20 May 2015


%\cite{Halpern:1983jt}
\bibitem{Halpern:1983jt} 
  M.~B.~Halpern,
\emph{Constrained quenched master field for continuum QCD},
  Nucl.\ Phys.\ B {\bf 228} (1983) 173.
  %%CITATION = NUPHA,B228,173;%%
  %24 citations counted in INSPIRE as of 20 May 2015

%\cite{Batrouni:1985qr}
\bibitem{Batrouni:1985qr}
  G.~G.~Batrouni, H.~Kawai and P.~Rossi,
\emph{Coordinate independent formulation of the Langevin equation},
  J.\ Math.\ Phys.\  {\bf 27} (1986) 1646.
  %%CITATION = JMAPA,27,1646;%%
  %4 citations counted in INSPIRE as of 07 Aug 2015

%% %\cite{Aarts:2008rr}
%% \bibitem{Aarts:2008rr}
%%   G.~Aarts and I.~O.~Stamatescu,
%% \emph{Stochastic quantization at finite chemical potential},
%%   JHEP {\bf 0809} (2008) 018 
%%   [{\tt arXiv:0807.1597 [hep-lat]}].
%%   %%CITATION = ARXIV:0807.1597;%%
%%   %71 citations counted in INSPIRE as of 07 Aug 2015

%% %\cite{Sexty:2014dxa}
%% \bibitem{Sexty:2014dxa} 
%%   D.~Sexty,
%%   %``New algorithms for finite density QCD,''
%%   PoS LATTICE {\bf 2014}, 016 (2014)
%%   [arXiv:1410.8813 [hep-lat]].
%%   %%CITATION = ARXIV:1410.8813;%%
%%   %11 citations counted in INSPIRE as of 29 Jul 2015


\end{thebibliography}
\end{document}